\pdfoutput=1 
\RequirePackage{ifpdf}
\ifpdf 
\documentclass[pdftex]{sigma}
\else
\documentclass{sigma}
\fi

\def\C{\mathbb{C}}

\def\R{\mathbb{R}}
\def\Z{\mathbb{Z}}
\def\a{\alpha}
\def\p{\partial}

\begin{document}

\allowdisplaybreaks

\renewcommand{\thefootnote}{$\star$}

\renewcommand{\PaperNumber}{039}

\FirstPageHeading

\ShortArticleName{Intertwining Symmetry Algebras of Quantum Superintegrable Systems}

\ArticleName{Intertwining Symmetry Algebras
\\
of Quantum Superintegrable Systems\footnote{This paper is a contribution to the Proceedings of the VIIth Workshop ``Quantum Physics with Non-Hermitian Operators''
     (June 29 -- July 11, 2008, Benasque, Spain). The full collection
is available at
\href{http://www.emis.de/journals/SIGMA/PHHQP2008.html}{http://www.emis.de/journals/SIGMA/PHHQP2008.html}}}

\Author{Juan A. CALZADA~$^\dag$, Javier NEGRO~$^\ddag$ and Mariano A. DEL OLMO~$^\ddag$}

\AuthorNameForHeading{J.A. Calzada, J. Negro and M.A. del Olmo}

\Address{$^\dag$~Departamento de  Matem\'atica Aplicada,
Universidad de  Valladolid,
 E-47011, Valladolid,  Spain}
\EmailD{\href{mailto:juacal@eis.uva.es}{juacal@eis.uva.es}}

\Address{$^\ddag$~Departamento de  F\'{\i}sica  Te\'orica,
Universidad de  Valladolid,
 E-47011, Valladolid,  Spain}
\EmailD{\href{mailto:jnegro@fta.uva.es}{jnegro@fta.uva.es}, \href{mailto:olmo@fta.uva.es}{olmo@fta.uva.es}}

\ArticleDates{Received November 14, 2008, in f\/inal form March 18,
2009; Published online April 01, 2009}

\Abstract{We present an algebraic study of a kind of quantum systems belonging to a family of superintegrable Hamiltonian systems in terms of shape-invariant intertwinig operators, that span pairs of Lie algebras like $(su(n), so(2n))$ or $(su(p,q), so(2p,2q))$.
The  eigenstates of the associated Hamiltonian hierarchies belong to
unitary representations of these algebras.
It is shown that these intertwining operators, related with separable
coordinates for the system, are very useful to determine eigenvalues and eigenfunctions of the Hamiltonians in the hierarchy. An study of the corresponding superintegrable classical systems is also included for the sake of completness.}


\Classification{17B80; 81R12; 81R15}

\section{Introduction}

It is well known that a Hamiltonian system (HS) in a conf\/iguration space of dimension $n$, is said to be integrable if there are $n$ constants of motion, including the Hamiltonian $H$, which are
independent and in involution.
If the systems has $0<k\leq n-1$  additional constants of motion
then, it is called superintegrable.
The physical system is said to be maximally superintegrable
if there exist  $2 n  -  1$ invariants well def\/ined in phase-space.
The superintegrable Hamiltonian systems (SHS) share nice properties. For instance, they admit  separation of variables  in more than one coordinate system for the Hamilton--Jacobi equation in the classical case and for the  Schr\"odinger equation in the quantum case.
Let us mention also that the f\/inite classical trajectories are closed (periodic), while the discrete energy levels are degenerate in the quantum case.

There is a limited number of this kind of physical systems as can be found in the works by
Evans~\cite{evans}. More recently,   we quote
the deformed
algebra approach to superintegrabi\-lity by Daskaloyannis and collaborators~\cite{daskaloyannis,daskaloyannis01} and the
superintegrability in constant curvature conf\/iguration spaces
 \cite{pogosyan1, s1,s2}. Among a long list of contributions we can also mention two former references.
 In 1975 Lakshmanan and Eswaran \cite{lakshmanan_75} analyzed the isotropic oscillator on a~3-sphere and in 1979 motived by this work Higgs \cite{higgs_79} studied versions of the Coulomb potential and of the harmonic oscillator living in the $N$-dimensional sphere and having
 $SO(N+1)$ and $SU(N)$  symmetry in classical and in quantum mechanics, respectively.

Some  years ago a new  family of SHS, was constructed from a
group-theoretical method based on the symmetry reduction.
These systems  come, using  the Marsden--Weinstein
reduction~\cite{marsden-weinstein}, from free systems in $\C^{p,q}$ presenting an initial  $U(p,q)$-symmetry \cite{olmo93}
 \[
H= \frac{c}{4}g^{\bar{\mu}\nu}\bar{p}_{\mu}p_{\nu} \quad
 \stackrel{\rm MW\, reduction}\longrightarrow \quad
H^r= \frac{c}{4}g^{\mu\nu}p_{s^\mu}p_{s^\nu} + V(s),
\]
where the bar stands for the complex conjugate and $V(s)$ is a potential in terms of the real coordinates ($s^\mu$). These SHS are living in   conf\/iguration spaces
of constant curvature ($SO(p,q)$-homogeneous spaces).

\looseness=1
Although the quantum version of these systems is well known
and can be exhaustively  studied in all its aspects
with standard procedures \cite{olmo93,olmo96,olmo97,olmo99a,olmo99b}, we present here a new perspective  based on intertwining operators (IO), a form of Darboux transformations \cite{darboux}, that will allow us to study them from an algebraic point of view.   The associated IO's close Lie algeb\-ras that take into account the symmetry properties of the systems and permit to describe these SHS in terms of representations of  such ``intertwining symmetry'' algebras (or ``dynamical algeb\-ras''~\cite{iachello}).

The intertwining operators are f\/irst order dif\/ferential operators,
$A$,
connecting dif\/ferent Hamiltonians, $H$, $H'$, in the same hierarchy, i.e.,
$A H=H' A$.
In the cases under study it is obtained a complete set of such~IO's,
in the sense that any of the Hamiltonians of the hierarchy can be
expressed in terms of these operators.
As we will see later, the IO's are associated to systems of separable coordinates for the Hamiltonians.
The study of the IO's associated to integrable Hamiltonians
has been made, for instance,  in~\cite{kuru01,kuru02,olmo96a}
and, following this line of research, we will supply here other
non-trivial applications by means of the above mentioned family of SHS.
From the perspective of the IO's,
we present a natural extension to higher dimensions of the intertwining (Darboux) transformations
of the Schr\"odinger equation for one-dimensional quantum systems~\cite{infeld-hull}. When a system of separable coordinates is used, any Hamiltonian of this family of SHS gives rise to a coupled set of  $n$-dif\/ferential equations, which can be factorized one by one.

\looseness=1
In principle, we will present two particular SHS that we denote $u(3)$-system \cite{olmo06} and  $u(2,1)$-system \cite{olmo08}, but the generalization to higher $u(p,q)$-systems is evident.   They are living in conf\/iguration spaces
of constant curvature ($SO(3)$ and $SO(2,1)$-homogeneous spaces, respectively): a 2D sphere and a 2D hyperboloid of two-sheets.
By extending  well known methods in one-dimension to higher-dimensional systems we obtain a wide set of IO's closing the
dynamical Lie algebras $u(3)$ or $u(2,1)$.
These initial intertwining symmetry algebras can be enlarged by considering discrete symmetry operators obtaining, respectively, the $so(6)$ and $so(4,2)$ Lie algebras of IO's.
This  approach  gives a simple explanation of the main features of
these physical systems. For instance, it allows us to characterize
the discrete spectrum and the corresponding eigenfunctions of the system by means of  (f\/inite/inf\/inite) irreducible unitary representations (IUR) of the (compact/non-compact)  intertwining symmetry algebras.
We can compute the ground state
and characterize the  representation space
of the wave-functions which share the same energy.
The organization of the paper is as follows. In Section~\ref{SuperintegrableSU(p,q)} we introduce
the classical superintegrable Hamiltonian family under consideration.  In Section~\ref{sqsystems}  we focus on the
quantum systems and  show  how to
build the IO's connecting hierarchies of these kind of Hamiltonians.
It is seen that these operators
close a $su(2,1)$ or a $su(3)$ Lie algebra.
The Hamiltonians are related to the second order
Casimirs of such  algebras, while the discrete spectrum of the
Hamiltonians is related to their IUR's. Next,  a broader class of IO's is
def\/ined leading to the $so(4,2)$ or $so(6)$ Lie algebras,
and it is shown
how this new structure helps us to understand better the
Hamiltonians in the new hierarchies. Finally, some remarks and
conclusions in Section~\ref{section4} will end the paper.

\section[Superintegrable $SU(p,q)$-Hamiltonian systems]{Superintegrable $\boldsymbol{SU(p,q)}$-Hamiltonian systems}
\label{SuperintegrableSU(p,q)}

Let us consider the free  Hamiltonian
\begin{gather}\label{hamiltonianolibre}
H=\,4\,g^{\mu\bar{\nu}} p_\mu \bar{p}_\nu, \qquad
\mu,\nu=0 ,\ldots,n= p+q-1,
\end{gather}
(by $p_\mu$ we denote the conjugate momenta) def\/ined in the conf\/iguration space
\[
 \frac{SU(p,q)}{SU(p-1,q)\times U(1)},
\]
 which is an Hermitian hyperbolic space with metric  $g_{{\mu}\nu}$ and coordinates
$y^\mu\in \mathbb{C}$ such that
\[
g_{\bar{\mu}\nu} \bar{y}^\mu y^\nu= 1 .
\]
The geometry and properties of this kind of spaces  are described in~\cite{kobayashi-nomizu} and \cite{olmo97}.

Using a maximal Abelian subalgebra (MASA) of  $su(p,q)$ \cite{olmo90}  the reduction procedure allows us to obtain a reduced Hamiltonian, which is not free,
lying in the corresponding reduced space, a homogeneous $SO(p,q)$-space \cite{olmo93,olmo96}
\[
 H= \frac{1}{2} g^{\mu \nu} p_{s^\mu} p_{s^\nu} + V(s) ,
\]
where $V(s)$ is a potential depending on the real coordinates
$s^\mu$ satisfying
$g_{{\mu}\nu} {s}^\mu s^\nu= 1$.

The set of complex coordinates $y^\mu$ after
the reduction procedure becomes a set of ignorable variables
$x^\mu$  and the actual real coordinates $s^\mu$.
A way to implement the symmetry reduction is as follows.
Let $Y_\mu$, $\mu= 0,\ldots,n$, be a basis of the considered MASA of
$u(p,q)$ constituted only by pure imaginary matrices (this is a basic hypothesis in the reduction procedure). Then the relation between
old $(y^\mu)$  and  new coordinates $(x^\mu,s^\mu)$ is
\[
    y^\mu= B(x)^{\mu}_{\nu} s^\nu ,\qquad
    B(x)=\exp\left(x^\mu Y_\mu\right).
 \]
The fact that  the $(x^\mu)$  are the parameters of the transformation associated to the  MASA of~$u(p,q)$ used in the
reduction,  assures the ignorability of the $x$ coordinates (in other words, the vector
f\/ields corresponding to the MASA are straightened out in these
coordinates).
The Jacobian matrix, $J$, corresponding to the coordinate transformation
($(y,\bar{y})  \to    (x,s)$) is given explicitly~by
\[
    J= \frac{\partial(y ,\bar{y})}{\partial (x ,s)}=
    \begin{pmatrix}
        A & B \\
        \bar{A} & \bar{B}
    \end{pmatrix} ,
\]
where
\[
    A_{ \nu}^{\mu} = \frac{\partial y^\mu}{\partial
    x^\nu}= \left(Y_\nu\right)^{\mu}_{ \rho} y^\rho .
\]
The expression of the Hamiltonian \eqref{hamiltonianolibre} in the new coordinates  $s$ is
\[
    H= c  \left(\frac{1}{2} g^{\mu \nu}p_\mu  p_\nu +
    V(s)\right) ,\qquad
 V(s)= p_{x}^{T} (A^{\dagger} K A)^{-1} p_x ,
 \]
where $p_x$ are the constant momenta  associated to the ignorable
coordinates $x$ and $K$ is the matrix def\/ined by the metric $g$.

\subsection[A classical superintegrable $u(3)$-Hamiltonian]{A classical superintegrable $\boldsymbol{u(3)}$-Hamiltonian}
\label{u(3)Hamiltonian}

To obtain the classical superintegrable Hamiltonian associated to  $su(3)$, using the
reduction procedure sketched before, we  proceed as follows:
let us consider the  basis of $su(3)$  determined by   $3 \times 3$ matrices
$X_1 ,\ldots,X_8$, whose explicit form,  using the metric  $K=\text{diag}(1,1,1)$, is
\begin{alignat*}{4}
    &X_1 = \begin{pmatrix} i & 0 & 0\\ 0 &-i & 0\\ 0 & 0 &0\end{pmatrix}, &\quad
    &X_2 = \begin{pmatrix} 0 & 0 & 0\\ 0 &i & 0\\ 0 & 0 &-i\end{pmatrix}, &\quad
    &X_3 = \begin{pmatrix} 0 & 1 & 0\\ -1 &0 & 0\\ 0 & 0 &0
    \end{pmatrix}, &\quad
    &X_4 = \begin{pmatrix} 0 & i & 0\\ i &0 & 0\\ 0 &
    0&0\end{pmatrix}, \\
    &X_5 = \begin{pmatrix} 0 & 0 & 1\\ 0 &0 & 0\\ -1 & 0 &0
    \end{pmatrix}, &\quad
   & X_6 = \begin{pmatrix} 0 & 0 & i\\ 0 &0 & 0\\ i & 0 &0\end{pmatrix}, &\quad
    &X_7 = \begin{pmatrix} 0 & 0 & 0\\ 0 &0 & 1\\ 0 & -1 &0\end{pmatrix}, &\quad
    &X_8 = \begin{pmatrix} 0 & 0 & 0\\ 0 &0 & i\\ 0 & i&0\end{pmatrix}.
\end{alignat*}
There is only one MASA for  $su(3)$:  the Cartan subalgebra,
generated by the matrices
\[
{\rm diag}(i,-i,0),\qquad {\rm diag}(0,i,-i).
\]
So, we can generate only one $su(3)$-Hamiltonian system.
In order to  facilitate the computations we  shall use the following basis for the corresponding MASA in $u(3)$
\begin{gather}\label{ccsu3}
    Y_0= {\rm diag}(i,0,0),\qquad
      Y_1= {\rm diag}(0,i,0),\qquad
    Y_2= {\rm diag}(0,0,i).
 \end{gather}
The actual real coordinates $s$ are related to the complex coordinates $y$ by
 \[
 y_\mu= s_\mu e^{i x_\mu}, \qquad  \mu=0,1,2 ,
 \]
and the Hamiltonian can be written as
\begin{gather}\label{hamiltonianosu3}
    H = \frac{1}{2} \left(p_{0}^2 + p_{1}^2+p_{2}^2\right) + V(s) ,\qquad
    V(s)= \frac{m_{0}^2}{s_{0}^2} +
    \frac{m_{1}^2}{s_{1}^2} +\frac{m_{2}^2}{s_{2}^2}  ,
\end{gather}
which lies in the 2-sphere  $ (s_0)^2+  (s_1)^2 +  (s_2)^2 = 1$, with
$m_{0}, m_{1},m_{2}\in \R$.
The system is superintegrable since there exist three invariants of motion
\begin{gather*}
  R_{\mu\nu}=  (s_\mu p_\nu - s_\nu p_\mu)^2 +
  \left(m_\mu \frac{s_\nu}{s_\mu} +
        m_\nu  \frac{s_\mu}{s_\nu} \right)^2 ,\qquad \mu<\nu, \quad  \mu=0,1,\quad \nu=1,2 .
\end{gather*}
The constants of motion $R_{\mu\nu}$ can be written in terms of the basis of $su(3)$ (in the realization as function of $s_\mu$ and $p_\mu$)
\[
Q_1\equiv R_{01}=X_3^2+X_4^2,\qquad
Q_2\equiv R_{02}=X_5^2+X_6^2,\qquad
Q_3\equiv R_{12}=X_7^2+X_8^2,
\]
and the sum of these invariants is the Hamiltonian
\eqref{hamiltonianosu3}   up to an additive constant
\begin{gather*}
H = Q_1+Q_2+Q_3 + \text{cnt}.
\end{gather*}
The quadratic Casimir of $su(3)$ can be also written in terms of the constants of motion  and the second order operators in the enveloping algebra of the compact Cartan subalgebra of $su(3)$
\[
{\cal C}_{su(3)}=  3 Q_1+3 Q_2+3 Q_3+4 X_1^2 +2 [X_1,X_2]_+ +4 X_2^2
.\]
The Hamiltonian is in involution with all the three constants of motion, i.e.\ $[H,Q_i]=0$, $ i=1,2,3$. However, the  $Q_i$'s do not commute among them
\begin{gather*}
[Q_1,Q_2]=[Q_3,Q_1]=[Q_2,Q_3]
\\
\phantom{[Q_1,Q_2]}{}  =-[X_3, [X_5,X_7]_+]_+ -[X_3,[X_6,X_8]_+]_+ +[X_4,[X_5,X_8]_+]_+
-[X_4,[X_6,X_7]_+]_+ .
\end{gather*}
So, the system \eqref{hamiltonianosu3} is superintegrable.

\subsubsection[The Hamilton-Jacobi equation for the $u(3)$-system]{The Hamilton--Jacobi equation for the $\boldsymbol{u(3)}$-system}\label{hamiltonjacobiequ(3)}

The  solutions of the motion problem for this system can be obtained  solving
 the corresponding Hamilton--Jacobi (HJ)
equation  in an appropriate coordinate system, such that the HJ
equation separates into a system of ordinary dif\/ferential equations.

The 2-sphere
can be  parametrized  on spherical coordinates
 $(\phi_1,\phi_2)$ around the $s_2$ axis    by
\[
   s_0 = \cos \phi_2 \cos \phi_1,\qquad
   s_1 = \cos \phi_2  \sin \phi_1,\qquad
   s_2  = \sin\phi_2 ,
\]
   where $\phi_1\in [0,2 \pi)$ and  $\phi_2\in [\pi/2,3 \pi/2]$.
Then, the Hamiltonian  \eqref{hamiltonianosu3} is rewritten as
\[
    H= \frac{1}{2} \left(p_{\phi_2}^2 +
        \frac{p_{\phi_1}^2}{\cos^2\phi_2}\right)+ \frac{1}{\cos^2
        \phi_2} \left(\frac{m_{0}^2}{\cos^2\phi_1} +
        \frac{m_{1}^2}{\sin^2 \phi_1} \right) + \frac{m_{2}^2}{\sin^2
        \phi_2}  .
\]
The potential is periodic and has singularities
along the coordinate lines
$\phi_1=0,\pi/2,\pi, 3\pi/2$  and
$\phi_2=\pi/2,3\pi/2$, and
there is  a unique minimum inside each
domain of  regularity.

The invariants $Q_i$  can be rewritten in spherical coordinates taking the explicit form
\begin{gather*}
    Q_1= \frac{1}{2} p_{\phi_1}^2 + \frac{m_{0}^2} \cos^2\phi_1  +
        \frac{m_{1}2} \sin^2\phi_1 ,\\
   Q_2= \tan^2\phi_2 \left(\frac{1}{2}p_{\phi_1}^2
         \sin^2\phi_1 + \frac{m_{0}^2} \cos^2\phi_1 \right) +
        \cos^2\phi_1 \left(\frac{1}{2} p_{\phi_2}^2 +
        \frac{m_{2}^2} \tan^2\phi_2 \right) \\
\phantom{Q_2=}{} +   \frac{1}{2} p_{\phi_1} p_{\phi_2} \sin 2 \phi_1 \tan \phi_2 ,\\
  Q_3=  \tan^2\phi_2 \left(\frac{1}{2}p_{\phi_1}^2
         \cos^2\phi_1 + \frac{m_{1}^2} \sin^2\phi_1 \right) +
        \sin^2\phi_1 \left(\frac{1}{2} p_{\phi_2}^2 +
        \frac{m_{2}^2} \tan^2\phi_2 \right)\\
\phantom{Q_3=}{} - \frac{1}{2} p_{\phi_1} p_{\phi_2} \sin 2 \phi_1 \tan \phi_2 .
\end{gather*}

Now, the HJ equation takes the form
\[
{1\over 2} \left( {\partial S\over \partial \phi_2}\right)^2
 +  {m_2^2\over\sin^2\phi_2}+
{1\over \cos^2\phi_2}\left({1\over 2}  \left( {\partial S\over
\partial \phi_1}\right)^2
+{m_0^2\over \cos^2\phi_1} + {m_1^2\over
\sin^2\phi_1}\right)=E .
\]
It separates into two ordinary
dif\/ferential equations taking into account that the solution of the HJ equation can be written as
$S(\phi_1,\phi_2)=S_1(\phi_1)+ S_2(\phi_2)-Et$. Thus,
\begin{gather*}
{1\over 2} \left( {\partial
S_1\over \partial
\phi_1}\right)^2+{m_0^2\over
\cos^2\phi_1} + {m_1^2\over
\sin^2\phi_1}   =   \alpha_1,\label{hjsu3}\\
{1\over 2} \left( {\partial S_2\over \partial
\phi_2}\right)^2 +  {m_2^2\over
\sin^2\phi_2}+  {\alpha_1\over \cos^2\phi_2}
  =   \alpha_2 ,
\end{gather*}
where $\alpha_2=E$ and $\alpha_1$ are the
separation constants (which are positive).
Each one of these two equations is formaly similar to
those  of the corresponding one-dimensional problem~\cite{olmo99a}.
The solutions of both HJ equations are easily computed
and can be found as particular cases in~\cite{olmo97}.
Notice that all the orbits in a
neighborhood of a critical point (center)
are closed and, hence, the corresponding
trajectories are periodic.

The explicit solutions, when we restrict us to the domain
$0<\phi_1,\phi_2<\pi/2$, are
\begin{gather*}
\cos^2\phi_2   =   {1\over 2E}\left[
b_2+\sqrt{b_2^2-4\alpha_1 E}\cos
2\sqrt{2E}t\right], \\
\cos^2\phi_1   =   {1\over 2\alpha_1}\Bigg[
b_1+{1\over\cos^2\phi_2}
\left[{b_1^2-4\alpha_1 m_0^2\over
b_2^2-4\alpha_1E}\right]^{1/2}\Big(
(b_2\cos^2\phi_2-2\alpha_1)\sin
2\sqrt{2\alpha_1}\beta_1
\nonumber\\
   \phantom{\cos^2\phi_1   =}{}
+2 \sqrt{\alpha_1}\big[(b_2
-E\cos^2\phi_2)\cos^2\phi_2-\alpha_1\big]^{1/2}
\cos 2\sqrt{2\alpha_1}\beta_1\Big)\Bigg] ,
\end{gather*}
where $b_1=\alpha_1+m_0^2-m_1^2$ and
$b_2=E+\alpha_1-m_2^2$.
Inside the domain the minimum for the potential is  at the point
$(\phi_1=\arctan\sqrt{m_1/m_0},\;\phi_2=\arctan  \sqrt{m_2/(m_0+m_1))}$, and its value  is
$V_{\min}=(m_0+m_1+m_2)^2$. Hence, the energy
$E$ is  bounded from below
$E\ge (m_0+m_1+m_2)^2$.

\subsection[A classical superintegrable $u(2,1)$-Hamiltonian]{A classical superintegrable $\boldsymbol{u(2,1)}$-Hamiltonian}
\label{u(2,1)Hamiltonian}

In a similar way to the preceding case $u(3)$ of Section~\ref{u(3)Hamiltonian}, it is enough to f\/ind an appropriate  basis of $u(2,1)$, for instance \begin{alignat*}{4}
    &X_1 = \begin{pmatrix} i & 0 & 0\\ 0 &-i & 0\\ 0 & 0 &0\end{pmatrix}, &\quad
    &X_2 = \begin{pmatrix} 0 & 0 & 0\\ 0 &i & 0\\ 0 & 0 &-i\end{pmatrix}, &\quad
    &X_3 = \begin{pmatrix} 0 & 1 & 0\\ -1 &0 & 0\\ 0 & 0 &0
    \end{pmatrix}, &\quad
    &X_4 = \begin{pmatrix} 0 & i & 0\\ i &0 & 0\\ 0 &
    0&0\end{pmatrix}, \\
    &X_5 = \begin{pmatrix} 0 & 0 & 1\\ 0 &0 & 0\\ 1 & 0 &0
    \end{pmatrix}, &\quad
    &X_6 = \begin{pmatrix} 0 & 0 & i\\ 0 &0 & 0\\ -i & 0 &0\end{pmatrix}, &\quad
    &X_7 = \begin{pmatrix} 0 & 0 & 0\\ 0 &0 & 1\\ 0 & 1 &0\end{pmatrix}, &\quad
    &X_8 = \begin{pmatrix} 0 & 0 & 0\\ 0 &0 & i\\ 0 &- i&0\end{pmatrix},
\end{alignat*}
and to follow the same procedure. However, the Lie algebra $su(2,1)$ has four MASAS~\cite{olmo90}: the compact Cartan subalgebra like   $u(3)$, the  noncompact Cartan subalgebra, the orthogonally decomposable subalgebra and the nilpotent subalgebra. For our purposes in this work we will only consider the symmetry reduction by the compact Cartan subalgebra, although we could generate other three SHS with the remaining MASAs. It is also possible to use  the same matrices of $u(3)$ \eqref{ccsu3} to build up a basis of the compact Cartan subalgebra.

 We f\/ind the following reduced Hamiltonian
\begin{gather}\label{hamiltonianosu21}
    H = \frac{1}{2}  c \left(-p_{0}^2 - p_{1}^2+p_{2}^2\right) +\frac{m_{0}^2}{s_{0}^2} +
    \frac{m_{1}^2}{s_{1}^2} -\frac{m_{2}^2}{s_{2}^2}  ,
\end{gather}
lying in the 2-dimensional two-sheet hyperboloid
$-s_{0}^2-s_{1}^2+s_{2}^2=1$ and with $c$ a constant. The potential constants, $m_i$, can be chosen non-negative real numbers.

Parametrizing  the two-sheet hyperboloid by  using an   `analogue' of the spherical coordinates
\begin{gather}\label{coordinatesu21}
s_{0}= \sinh\xi  \cos \theta ,\qquad
s_{1}= \sinh\xi  \sin \theta ,\qquad
s_{2}= \cosh \xi  ,
\end{gather}
with $0\leq\theta<\pi /2$ and  $0\leq\xi<\infty$,
the Hamiltonian can be rewritten if $c=-1$ as
\[
    H= \frac{1}{2} \left(p_{\xi}^2 +
        \frac{p_{\theta}^2}{\sinh^2\xi}\right) +
 \frac{1}{\sinh^2 \xi} \left(\frac{m_{0}^2}{\cos^2\theta} +
        \frac{m_{1}^2}{\sin^2 \theta} \right) - \frac{m_{2}^2}{\cosh^2
        \xi}  .
\]
The potential is regular inside the domain of the variables and there is   a saddle point for the values $\theta=\arctan \sqrt{m_1/m_0}$ and
$  \xi=\arg\tanh\sqrt{m_2(m_0+m_1)})$  if $m_0+m_1>m_2$.

The quadratic constants of motion in terms of the enveloping algebra of $su(2,1)$ are
\[
Q_1=X_3^2+X_4^2,\qquad
Q_2 =X_5^2+X_6^2,\qquad
Q_3 =X_7^2+X_8^2,
\]
and the sum of these invariants gives also the Hamiltonian up to an additive
constant
\[
H = -Q_1+Q_2+Q_3 + \text{cnt}.
\]
The quadratic Casimir of $su(2,1)$  is
\[
{\cal C}_{su(2,1)}=  3 Q_1- 3 Q_2- 3 Q_3+4 X_1^2 +2 [X_1,X_2]_+ + 4 X_2^2 .
\]
The Hamiltonian is in involution with all the three constants of motion that do not commute among  themselves
 \begin{gather*}
[Q_1,Q_2] = [Q_3,Q_1]=[Q_3,Q_2]
\\
\phantom{[Q_1,Q_2]}{} =  -[X_3, [X_5,X_7]_+]_+ -[X_3,[X_6,X_8]_+]_+
+[X_4,[X_5,X_8]_+]_+ -[X_4,[X_6,X_7]_+]_+ .
\end{gather*}
The explicit form of the invariants of motion in terms of the coordinates $\xi$ and $\theta$ is
 \begin{gather*}
    Q_1= \frac{1}{2} p_{\theta}^2 + \frac{m_{0}^2}{\cos^2\theta} +
        \frac{m_{1}^2}{\sin^2\theta},\\
   Q_2= \coth^2\xi \left(\frac{1}{2}p_{\theta}^2
         \sin^2\theta + \frac{m_{0}^2}{\cos^2\theta}\right) +
        \cos^2\theta\,\left(\frac{1}{2} p_{\xi}^2 +
        \frac{m_{2}^2}{\coth^2\xi}\right)    +   \frac{1}{2} p_{\theta} p_{\xi} \sin 2 \theta \coth \xi ,\\
  Q_3=  \coth^2\xi \left(\frac{1}{2}p_{\theta}^2
         \cos^2\theta + \frac{m_{1}^2}{\sin^2\theta}\right) +
        \sin^2\theta \left(\frac{1}{2} p_{\xi}^2 +
        \frac{m_{2}^2}{\coth^2\xi}\right)   - \frac{1}{2} p_{\theta} p_{\xi} \sin 2 \theta \coth \xi .
\end{gather*}

\section{Superintegrable quantum   systems}
\label{sqsystems}

In the previous Section~\ref{SuperintegrableSU(p,q)} we have described some classical superintegrable systems. Now we will study their quantum versions.
In order to construct the quantum version of both systems, let us proceed in the following way. By inspection of the classical Hamiltonians~\eqref{hamiltonianosu3} and~\eqref{hamiltonianosu21} and their constants of motions we can relate the terms like $s_\mu  p_\nu \pm  s_\nu p_\mu$ with   generators of ``rotations'' when $p_\mu \to \partial_\mu$ in the plane $X_\mu X_\nu$. Moreover, since the Hamiltonian is the sum of the three constants of motion up to constants we can write a quantum Hamiltonian as a linear combination of $J_0^2$, $J_1^2$ and $J_2^2$, being $J_0$, $J_1$ and $J_2$ the inf\/initesimal generators of ``rotations'' in the plane  $X_1 X_2$, $X_0 X_2$ and $X_0 X_1$ around the axis $X_0$, $X_1$ and $X_2$, respectively. According to the signature of the metric the rotations will be compact or noncompact,   so, the generators will span~$so(3)$ or~$so(2,1)$  and for our purposes we will take a dif\/ferential realization of them. In other words, the Casimir  operator of $so(3 )$
($C_{so(3)}=J_{0}^2+J_{1}^2+J_{2}^2$) or of $so(2,1)$
($C_{so(2,1)}=J_{0}^2+J_{1}^2-J_{2}^2$)
gives the ``kinetic'' part of the corresponding Hamiltonian. In the following we will present a detailed study of the case related with
$su(2,1)$. The $su(3)$-case is simply sketched and the interested reader can  f\/ind more details in~\cite{olmo06}.

\subsection[Superintegrable quantum   $u(2,1)$-system]{Superintegrable quantum   $\boldsymbol{u(2,1)}$-system}
\label{su21sqsystems}

Let us  consider the Hamiltonian
\begin{gather}\label{hamiltoniansu(2,1)}
H_{\ell}=-J_{0}^2-J_{1}^2+J_{2}^2
+\frac{l_{0}^2-\frac{1}{4}}{s_{0}^2}
+\frac{l_{1}^2-\frac{1}{4}}{s_{1}^2}-
\frac{l_{2}^2-\frac{1}{4}}{s_{2}^2} ,
\end{gather}
which conf\/iguration space  is the 2-dimensional two-sheet hyperboloid
$-s_{0}^2-s_{1}^2+s_{2}^2=1$,  with
$\ell=(l_{0}, l_{1}, l_{2}) \in \mathbb{R}^{3}$ (and $2  m=1$).
The dif\/ferential operators
\[
J_0=s_1\partial_2+ s_2\partial_1,\qquad
J_1=s_2\partial_0+ s_0\partial_2,\qquad
J_2=s_0\partial_1- s_1\partial_0 ,
\]
constitute a realization of  $so(2,1)$
with Lie commutators
\[
[J_{0},J_{1}]=-J_{2}, \qquad
[J_{2},J_{0}]=J_{1},\qquad
[J_{1},J_{2}]=J_{0}.
\]
Using  coordinates \eqref{coordinatesu21}
the explicit expressions of the inf\/initesimal generators are
\[
J_{0}=\sin{\theta} \partial_{\xi}+\cos \theta   \coth\xi
 \partial_{\theta} ,\qquad
J_{1}=\cos \theta  \partial_{\xi}-\sin \theta   \coth\xi
 \partial_{\theta} ,\qquad
J_{2}=\partial_{\theta}  .
\]
They are anti-Hermitian operators inside the space of square-integrable functions with  invariant measure
$d\mu(\theta,\xi)= \sinh\xi\, d\theta d\xi$.

In these coordinates the Hamiltonian  $H_{\ell}$
\eqref{hamiltoniansu(2,1)}
 has the expression
\[
H_{\ell}=-\partial_{\xi}^2-{\coth\xi}\,\partial_{\xi}
-\frac{l_{2}^2-\frac{1}{4}}{\cosh^{2}\xi}
+\frac{1}{\sinh^{2}\xi}
  \left[-\partial_{\theta}^2+\frac{l_{1}^2
-\frac{1}{4}}{\sin^{2}\theta}+\frac{l_{0}^2
-\frac{1}{4}}{\cos^{2}\theta}\right].
\]
It  can be separated in the variables $\xi$ and
$\theta$, by choosing its eigenfunctions $\Phi_{\ell}$
 in the form
$\Phi_{\ell}(\theta,\xi)=f(\theta)\,g(\xi)$,
 obtaining a pair of  separated equations
\begin{gather}\label{hamiltoniansu(2,1)a}
H_{l_{0},l_{1}}^{\theta} f(\theta)\equiv
\left[-\partial_{\theta}^2+\frac{l_{1}^2-\frac{1}{4}}{\sin^{2}\theta}
+\frac{l_{0}^2-\frac{1}{4}}{\cos^{2}\theta}\right]f(\theta)
=\alpha\,f(\theta) ,
\\
\left[-\partial_{\xi}^2-{\coth\xi} \partial_{\xi}
-\frac{l_{2}^2-\frac{1}{4}}{\cosh^{2}\xi}
+\frac{\alpha}{\sinh^{2}\xi}\right]g(\xi)=E g(\xi) ,\nonumber
 \end{gather}
where $\alpha>0$ is a separation constant.

\subsubsection[A complete  set of intertwining operators for $H_{\ell}$]{A complete  set of intertwining operators for $\boldsymbol{H_{\ell}}$} \label{completesetofintertwiningoperators}

The one-dimensional Hamiltonian
$H_{l_{0},l_{1}}^{\theta}$ \eqref{hamiltoniansu(2,1)a} can be factorized  as a product of f\/irst order opera\-tors~$A^\pm$ and a constant $\lambda_{(l_0,l_1)}$
\begin{gather}
H_{(l_0,l_1)}^{\theta} = A^{+}_{(l_0,l_1)} A^{-}_{(l_0,l_1)} + \lambda_{(l_0,l_1)},\nonumber\\
A_{l_{0},l_{1}}^{\pm}=\pm\partial_{\theta}-(l_{0}+1/2) {\tan{\theta}}
+(l_{1}+1/2) {\cot{\theta}} ,  \label{operadoresA}\\
\lambda_{(l_0,l_1)} =  (l_0 + l_1+1 )^2 .\nonumber
\end{gather}
The fundamental relation between contiguous couples of operators $A^\pm$,
\[
H_{(l_0,l_1)}^{\theta} =
A^{+}_{(l_0,l_1)} A^{-}_{(l_0,l_1)} + \lambda_{(l_0
,l_1)} =
A^{-}_{(l_0+1,l_1+1)} A^{+}_{(l_0+1,l_1+1)} +
    \lambda_{(l_0+1,l_1+1)} ,
\]
 allows us to construct a hierarchy of Hamiltonians
\[
 \ldots , \ H_{{l_{0}-1},{l_{1}-1}}^{\theta}, \ H_{{l_{0}},{l_{1}}}^{\theta}, \ H_{{l_{0}+1},{l_{1}+1}}^{\theta}, \
\ldots, \ H_{{l_{0}+n},{l_{1}+n}}^{\theta}, \ \ldots ,
\]
which satisfy the  recurrence relations
\begin{gather*}
A^{-}_{(l_0,l_1)} H_{(l_0,l_1)}^{\theta}=
H_{(l_0{+}1,l_1{+}1)}^{\theta} A^{-}_{(l_0,l_1)}, \qquad
A^{+}_{(l_0,l_1)} H_{(l_0{+}1,l_1{+}1)}^{\theta}=H_{(l_0,l_1)}^{\theta} A^{+}_{(l_0,l_1)}.
\end{gather*}
From the above relations we see that  the operators
$A_{(l_0,l_1)}^\pm$ act  as shape invariant intertwining operators and also that  $A^{-}_{(l_0,l_1)}$ transforms  eigenfunctions  of
$H_{(l_0,l_1)}^{\theta}$
into eigenfunctions of $H_{(l_0+1,l_1+1)}^{\theta}$, and
viceversa for $A^{+}_{(l_0,l_1)}$,
 in such a way that the original and the transformed eigenfunctions have the same eigenvalue.

Hence, once  the initial  values for $(l_0,l_1)$ have been  f\/ixed we can build up an  inf\/inite set of Hamiltonians
$\{H_{(l_0+n,l_1+n)}^{\theta} \}_{ n\in \Z}$
connected by  the set of operators
$  \{ A^{\pm}_{(l_0+n,l_1+n)} \}_{ n\in \Z}$ (Hamiltonian `hierarchy').
\subsubsection[The $u(2)$ 'dynamical' algebra]{The $\boldsymbol{u(2)}$ `dynamical' algebra}
\label{u2dynamicalalgebra}

We can def\/ine
free-index operators $\hat H^{\theta},\hat A^\pm,\hat A$ starting from the set of index-depending operators
$\{H_{(l_0+n,l_1+n)}^{\theta},A^\pm_{(l_0+n,l_1+_n)}\}_{n\in \Z}$. The free-index operators
act on the eigenfunctions $f_{(l_0+n,l_1+n)}$ of
$H_{(l_0+n,l_1+n)}^{\theta}$ as follows:
\begin{gather*}
\hat H^\theta f_{(l_0,l_1)} :=   H_{(l_0,l_1)}^{\theta} f_{(l_0,l_1)},
\\
\hat A^- f_{(l_0,l_1)} :=   \frac12 A^-_{(l_0,l_1)} f_{(l_0,l_1)},
\\
\hat A^+ f_{(l_0+1,l_1+1)} :=   \frac12 A^+_{(l_0,l_1)} f_{(l_0+1,l_1+1)},
\\
\hat A \,f_{(l_0,l_1)} :=   -\frac12(l_0 {+} l_1 ) f_{(l_0,l_1)}  .
\end{gather*}
With this convention the free-index operators close a $su(2)$-algebra
with commutators
\begin{gather}\label{commutatossu2}
[\hat A,\hat A^{\pm}] = \pm A^{\pm},\qquad
[\hat A^+,\hat A^-] =  2 \hat A ,
\end{gather}
and including  the operator
$D f_{(l_0,l_1)} := (l_0 -l_1)f_{(l_0,l_1)}$, that commutes with the other three ones, we obtain a~$u(2)$-algebra.

The fundamental states of some distinguished  Hamiltonians are in relation with the IUR's of~$su(2)$.
Thus, an eigenstate $f_{(l_0+n,l_1+n)}^0$ of
$H_{(l_0+n,l_1+n)}^{\theta}$ will be a fundamental (highest or lowest weight) vector if
\[
A^{-}  f_{(l_0+n,l_1+n)}^0 =
A^{-}_{(l_0+n,l_1+n)}  f_{(l_0+n,l_1+n)}^0 = 0 .
\]
The solution of this equation,
\begin{gather}\label{su3functions}
f_{(l_0+n,l_1+n)}^0(\theta_1)= N \cos^{l_0+1/2 + n}(\theta_1)
    \sin^{l_1+1/2+n}(\theta_1)
 \end{gather}
with $N$ a normalization constant and eigenvalue
\[
    E^0_{(l_0+n,l_1+n)} =  \lambda_{(l_0+n,l_1+n)}=
    (l_0 + l_1 +1+ 2 n)^2 ,
 \]
is also eigenfunction of $A$
\begin{gather}\label{eigenvalueA}
A   f_{(l_0+n,l_1+n)}^0  =
A_{(l_0+n,l_1+n)}  f_{(l_0+n,l_1+n)}^0 =
-\frac12(l_0+l_1+2n) f_{(l_0+n,l_1+n)}^0 .
\end{gather}
The functions \eqref{su3functions} are regular and square-integrable when $ l_0,l_1\geq -1/2$. From \eqref{eigenvalueA}
we can make the identification
\[
f_{(l_0+n,l_1+n)}^0 \simeq |j_n,-j_n\rangle,
\]
with $j_n=\frac{1}{2} (l_{0}+l_{1}+2n)$, $n=0,1,2,\dots$ The representation, $D^{j_n }$, f\/ixed by
$f_{(l_0+n,l_1+n)}^0$ will be a~IUR of~$su(2)$ of dimension
$2j_n+1=l_{0}+l_{1}+2n+1$ if
$l_0+l_1 \in \Z^+$, $n\in \Z^+$. The Hamilto\-nian~$H^\theta$ can be written in terms of the Casimir of $su(2)$, ${\cal C}= A^+A^- +A(A-1)$, as follows
\[
H^\theta =4({\cal C} +1/4).
\]
The other eigenstates in the representation $D^{j_n }$ are obtained applying recursively  $A^+$. Thus,
\[
  f_{(l_0,l_1)}^{n} = (A^+)^n f_{(l_0+n,l_1+n)}^0 =
A^{+}_{(l_0,l_1)} A^{+}_{(l_0+1,l_1+1)}\cdots
A^{+}_{(l_0{+}n{-}1,l_1{+}n{-}1)} f_{(l_0+n,l_1+n)}^0,
\]
and
\[
f_{(l_0,l_1)}^n\simeq |j_n,-j_n+n\rangle .
\]
The explicit form of $f_{(l_0,l_1)}^n$ is
\[
f_{(l_0,l_1)}^{n} =
     \sin^{l_1+1/2}(\phi_1) \cos^{l_0+1/2}(\phi_1)
       P_{m}^{(l_1,l_0)}[\cos(2 \phi_1)] ,
       \]
being $P_{m}$ the Jacobi polynomials, with eigenvalue
\[
E^n_{(l_0,l_1)}=(l_0 + l_1 +1+ 2\,n)^2,\qquad n\in\Z^+.
\]
Therefore, the eigenstates of the hierarchy
$\{H_{(l_0+n,l_1+n)}\}_{n\in \Z}$ when $l_0+l_1  \in \Z^+$ can be `organized' in IUR's of $su(2)$ (or of $u(2)$).
Notice that dif\/ferent fundamental states with values of~$l_0$ and~$l_1$,  such that    $j_0=(l_0+l_1)/2$ is f\/ixed, would lead to the same $j$-IUR of $su(2)$, but dif\/ferent $u(2)$-IUR's may correspond to states with the same energy (because $D f_{(l_0,l_1)}= (l_0-l_1) f_{(l_0,l_1)}$).
Hence, these results push us to f\/ind a larger  algebra of operators such that  all the eigenstates with the same energy  belong to only one of its IUR's.

Since  the IO's  $A^{\pm}_{l_{0},l_{1}}$ depend
only on the $\theta$-variable, they can act also as IO's of the
complete Hamiltonians $H_{\ell}$ \eqref{hamiltoniansu(2,1)}   and its global
eigenfunctions $\Phi_{\ell}$, leaving the parameter
$l_2$ unchanged
\[
A_{\ell'}^{-}H_{\ell'}=H_{\ell}A_{\ell'}^{-},\qquad
A_{\ell'}^{+}H_{\ell}=H_{\ell'}A_{\ell'}^{+},
\]
where $\ell=(l_{0},l_{1},l_2)$ and $\ell'=(l_{0}-1,l_{1}-1,l_2)$.
In this sense, many of the above relations can be straightforwardly
extended under this global point of view.

\subsubsection{Second set of pseudo-spherical coordinates}
\label{secondset}

A second coordinate set, obtained from the noncompact rotations around the axes $s_{2}$ and $s_{0}$ respectively, and that allows us to parametrize the hyperboloid
and separate the Hamiltonian is the following one
\[
s_{0}= \cosh\psi   \sinh\chi ,\qquad
s_{1}= \sinh\psi ,\qquad
s_{2}= \cosh\psi   \cosh\chi ,
\]
 with $-\infty <\psi<+\infty$ and $0 \leq \chi<+\infty$.
In these coordinates the $so(2,1)$-generators take the expressions
\[
J_{0}=- \tanh \psi   \sinh \chi  \partial_{\chi}
+ \cosh \chi  \partial_{\psi},\qquad
J_{1}=\partial_{\chi} ,\qquad
J_{2}= \sinh \chi  \partial_{\psi}- \tanh \psi
  \cosh \chi  \partial_{\chi} .
\]
The explicit expression of the Hamiltonian  is now
\[
H_{\ell}=-\partial_{\psi}^2- \tanh\psi  \partial_{\psi}
+\frac{l_{1}^2-\frac{1}{4}}{\sinh^{2}\psi}
+\frac{1}{\cosh^{2}\psi}
\left[-\partial_{\chi}^2+\frac{l_{0}^2
-\frac{1}{4}}{\sinh^{2}\chi}-
\frac{l_{2}^2-\frac{1}{4}}{\cosh^{2}\chi}\right] .
\]
It can be separated in the variables $\psi$ and $\chi$
considering the eigenfunctions   $H_\ell$  of the form
$\Phi(\chi,\psi)=f(\chi) g(\psi)$.
Hence, we obtain the  following two equations
\begin{gather}\label{hamiltoniansu(2,1)b}
H_{l_{0},l_{2}}^{\chi}f(\chi)\equiv
\left[-\partial_{\chi}^2+\frac{l_{0}^2
-\frac{1}{4}}{\sinh^{2}\chi}-\frac{l_{2}^2
-\frac{1}{4}}{\cosh^{2}\chi}\right]f(\chi)
=\alpha f(\chi) ,
\\
\left[-\partial_{\psi}^2
-{\tanh\psi} \partial_{\psi}+\frac{l_{1}^2
-\frac{1}{4}}{\sinh^{2}\psi}+\frac{\alpha}{\cosh^{2}\psi}\right]g(\psi)
=Eg(\psi),\nonumber
\end{gather}
with $\alpha$ a separation constant.

The Hamiltonian $H_{l_{0},l_{2}}^{\chi}$ \eqref{hamiltoniansu(2,1)b}
 can be factorized as a product of
f\/irst order operators $B^{\pm}$
\[
H_{l_{0},l_{2}}^{\chi}=
 B^{+}_{l_{0},l_{2}}B^{-}_{l_{0},l_{2}}
+\lambda_{l_{0},l_{2}}=
B^{-}_{l_{0}-1,l_{2}-1}B^{+}_{l_{0}-1,l_{2}-1}+\lambda_{l_{0}-1,l_{2}-1} ,
\]
being
 \begin{gather*}
B_{l_{0},l_{2}}^{\pm}=\pm\partial_{\chi}+(l_{2}+1/2)
\tanh{\chi}+(l_{0}+1/2)
  \coth \chi  ,\qquad
\lambda_{l_{0},l_{2}}=-(1+l_{0}+l_{2})^2 .
\end{gather*}
In this case the intertwining relations take the form
\[
B^{-}_{l_{0}-1,l_{2}-1}H_{l_{0}-1,l_{2}-1}^{\chi}=H_{l_{0},l_{2}}^{\chi}
B^{-}_{l_{0}-1,l_{2}-1},\qquad
B^{+}_{l_{0}-1,l_{2}-1}H_{l_{0},l_{2}}^{\chi}=
H_{l_{0}-1,l_{2}-1}^{\chi}B^{+}_{l_{0}-1,l_{2}-1} .
\]
Hence, the  operators $B^{\pm}$ connect eigenfunctions of
$H_{l_{0},l_{2}}^{\chi}$ in the following way
\[
B^{-}_{l_{0}-1,l_{2}-1}: \ f_{l_{0}-1,l_{2}-1}\rightarrow
f_{l_{0},l_{2}},\qquad
B^{+}_{l_{0}-1,l_{2}-1}: \ f_{l_{0},l_{2}}\rightarrow
f_{l_{0}-1,l_{2}-1} .
\]
The operators $B^{\pm}_{l_{0},l_{2}}$ can be also expressed in terms of $\xi$ and $\theta$
\[
B^{\pm}_{l_{0},l_{2}}=
\pm (\cos \theta  \partial_{\xi}-\sin \theta   \coth\xi
 \partial_{\theta} )+ (l_{2}+1/2)
 \tanh \xi \cos \theta +(l_{0}+1/2)  \coth \xi  \sec \theta  .
 \]
We also def\/ine new free-index operators
\[
\hat{B}^{-} f_{l_{0},l_{2}}:=\frac{1}{2} B_{l_{0},l_{2}}^{-}
 f_{l_{0},l_{2}},\qquad
\hat{B}^{+}\,f_{l_{0},l_{2}}:=\frac{1}{2} B_{l_{0},l_{2}}^{+}
 f_{l_{0},l_{2}},\qquad
\hat{B} f_{l_{0},l_{2}}:=-\frac{1}{2} (l_{0}+l_{2}) f_{l_{0},l_{2}},
\]
that close a $su(1,1)$ Lie algebra
\begin{gather}\label{commutatossu11}
[\hat{B}^{+}, \hat{B}^{-}]=-2\,\hat{B},\qquad
[\hat{B},\hat{B}^{\pm}]=\pm \hat{B}^{\pm}.
\end{gather}
Since  $su(1,1)$ is non-compact, its
IUR's are inf\/inite-dimensional.
In this case we are interested in the discrete series
having a fundamental state annihilated by the lowering operator
\[
B^{-}  f^0_{l_{0},l_{2}}=0 .
\]
The explicit
expression of these states is
\[
f^0_{l_{0},l_{2}}(\chi)
= N (\cosh \chi)^{l_2+1/2}(\sinh \chi)^{l_0+1/2} ,
\]
where $N$ is a normalization constant.
In order to have a regular
and square-integrable function we impose that
\[
l_0\geq -1/2,\quad -k_1\equiv l_0+l_2<-1   .
\]
Since
$
\hat{B} f^0_{l_{0},l_{2}}=
-\frac{1}{2} (l_{0}+l_{2}) f^0_{l_{0},l_{2}} ,
$
the lowest weight of this inf\/inite-dimensional  IUR of $su(1,1)$  is
characterized by
\[
j_1'= k_1/2>1/2.
\]

The IO's $\hat B^{\pm}$  can also be considered as intertwining operators  of the Hamiltonians $H_{\ell}$  linking their eigenfunctions
$\Phi_{\ell}$, similarly to the IO's $\hat A^{\pm}$, described before,  but now with $l_1$  remaining unchanged.

\subsubsection{Third set of pseudo-spherical coordinates}
\label{thirdset}

A third set of coordinates is obtained from the noncompact rotations around the axes~$s_{1}$ and~$s_{0}$, respectively. It  gives rise to the following parametrization of the hyperboloid
\[
s_{0}= \sinh\phi ,\qquad
s_{1}= \cosh\phi   \sinh\beta ,\qquad
s_{2}= \cosh\phi \cosh\beta ,
\]
with $0\leq \phi <+\infty$  and $-\infty< \beta <+\infty$.
The inf\/initesimal generators have the expressions
\[
J_{0}=\partial_{\beta},\qquad
J_{1}={\cosh{\beta}} \partial_{\phi}-{\tanh{\phi}}\,{\sinh{\beta}}
\;\partial_{\beta} ,\qquad
J_{2}=-{\sinh{\beta}} \partial_{\phi}
+{\tanh{\phi}}\,{\cosh{\beta}} \partial_{\beta} .
\]
Hence, the Hamiltonian now takes the form
\[
H_{\ell}=-\partial_{\phi}^2-{\tanh\phi} \partial_{\phi}
+\frac{l_{0}^2-\frac{1}{4}}{\sinh^{2}\phi}
+\frac{1}{\cosh^{2}\phi}
\left[-\partial_{\beta}^2+\frac{l_{1}^2
-\frac{1}{4}}{\sinh^{2}\beta}-\frac{l_{2}^2
-\frac{1}{4}}{\cosh^{2}\beta}\right] ,
\]
and it separates in the variables $\phi$, $\beta$ in terms of  its eigenfunctions
$\Phi(\beta,\phi)=f(\beta) g(\phi)$
\begin{gather*}
H_{l_{1},l_{2}}^{\beta}f(\beta)\equiv
\left[-\partial_{\beta}^2+\frac{l_{1}^2
-\frac{1}{4}}{\sinh^{2}\beta}-\frac{l_{2}^2
-\frac{1}{4}}{\cosh^{2}\beta}\right]f(\beta)=\alpha f(\beta) ,
\\
\left[-\partial_{\phi}^2
-{\tanh\phi} \partial_{\phi}+\frac{l_{0}^2-\frac{1}{4}}{\sinh^{2}\phi}
+\frac{\alpha}{\cosh^{2}\phi}\right]g(\phi)=
E\,g(\phi) ,\nonumber
\end{gather*}
with the separation constant $\alpha$.

The second order operator
$H_{l_{1},l_{2}}^{\beta}$ can be factorized as a product
of f\/irst order operators $C^\pm$
\[
H_{l_{1},l_{2}}^{\beta}={C}^{+}_{l_{1},l_{2}}{C}^{-}_{l_{1},l_{2}}+{\lambda}_{l_{1},l_{2}}
={C}^{-}_{l_{1}+1,l_{2}-1}{C}^{+}_{l_{1}+1,l_{2}-1}+
{\lambda}_{l_{1}+1,l_{2}-1},
\]
being
 \begin{gather*}
C_{l_{1},l_{2}}^{\pm}=
\pm\partial_{\beta}+(l_{2}+1/2)  \tanh \beta +(-l_{1}+1/2)
 \coth \beta ,\qquad
\lambda_{l_{1},l_{2}}=-(1-l_{1}+l_{2})^2 .
\end{gather*}
The operators ${C}^{\pm}_{l_{1},l_{2}}$ give rise to the intertwining relations
\[
{C}^{+}_{l_{1}+1,l_{2}-1}H_{l_{1},l_{2}}^{\beta}=
H_{l_{1}+1,l_{2}-1}^{\beta}C^{+}_{l_{1}+1,l_{2}-1}, \qquad
{C}^{-}_{l_{1}+1,l_{2}-1}H_{l_{1}+1,l_{2}-1}^{\beta}=
H_{l_{1},l_{2}}^{\beta}{C}^{-}_{l_{1}+1,l_{2}-1},
\]
which imply the connection among eigenfunctions
\[
{C}^{-}_{l_{1}+1,l_{2}-1}: \ f_{l_{1}+1,l_{2}-1}\rightarrow
f_{l_{1},l_{2}},\qquad
{C}^{+}_{l_{1}+1,l_{2}-1}: \ f_{l_{1},l_{2}}\rightarrow
f_{l_{1}+1,l_{2}-1} .
\]
The  IO's $C^{\pm}_{l_{1},l_{2}}$ can also be expressed in
terms of the f\/irst set of coordinates $(\xi, \theta)$
\[
C^{\pm}_{l_{1},l_{2}}=\pm (\sin \theta \partial_{\xi}+\cos \theta \coth\xi
 \partial_{\theta})+
(l_{2}+1/2)
 \tanh \xi \sin \theta +(-l_{1}+1/2)  \coth \xi  \csc \theta  .
\]
New free-index operators
are def\/ined as
\[
\hat{C}^{-} f_{l_{1},l_{2}}:=\frac{1}{2} {C}_{l_{1},l_{2}}^{-}
 f_{l_{1},l_{2}},\qquad
\hat{C}^{+} f_{l_{1},l_{2}}:=\frac{1}{2} {C}_{l_{1},l_{2}}^{+}
 f_{l_{1},l_{2}},\qquad
\hat{C} f_{l_{1},l_{2}}:=-\frac{1}{2} (l_{2}-l_{1})f_{l_{1},l_{2}},
\]
 satisfying  the commutation relations of the $su(1,1)$ algebra
\begin{gather}\label{commutatossu11a}
[\hat{C}^{-}, \hat{C}^{+}]=2\,\hat{C},\qquad
[\hat{C}, \hat{C}^{\pm}]=\pm \hat{C}^{\pm}.
\end{gather}
The fundamental state for the $su(1,1)$ representation,
 given by
$\hat {C}^{-} f^0_{l_{1},l_{2}}=0$,
 has the expression
\[
f^0_{l_{1},l_{2}}(\beta)=
N (\cosh \beta)^{l_2 +1/2} (\sinh \beta)^{-l_1 +1/2} ,
\]
with $N$  a normalization constant.
In order to get an IUR from
this eigenfunction, we impose it to be regular and normalizable, therefore
\[
l_1\leq 1/2,\qquad
-k_2\equiv l_2-l_1< -1  .
\]
The lowest weight of the IUR is given by
\[
j'_2= k_2/2>1/2,
\]
because in this case we have that
$\hat{C} f^0_{l_{1},l_{2}}=
-\frac{1}{2} (l_{2}-l_{1})f^0_{l_{1},l_{2}}$.

As in the other cases the IO's {$C^{\pm}$ can be considered
as connecting global Hamiltonians $H_{\ell}$} and their
eigenfunctions, having in mind that now the parameter {$l_0$ is
unaltered}.

\subsubsection{Algebraic structure of the intertwining operators}
\label{algebraicsu3}

If we consider together all the IO's
$\{\hat{A}^{\pm},\hat{A},\hat{B}^{\pm},
\hat{B},\hat{C}^{\pm},\hat{C}\}$
we f\/ind that they close a
$ {su(2,1)}$ Lie algebra, whose Lie commutators are displayed in
\eqref{commutatossu2}, \eqref{commutatossu11} and \eqref{commutatossu11a} together with the crossed commutators
\begin{alignat*}{4}
&[\hat{A}^{+}, \hat{B}^{+}]=0, \qquad&&
 [\hat{A}^{+},\hat{B}^{-}]=-\hat{C}^{-}, \qquad &&
 [\hat{A}^{+},\hat{B}]=-\frac{1}{2} \hat{A}^{+},& \\
&[\hat{A}^{+},\hat{C}^{+}]=\hat{B}^{+}, \qquad&&
 [\hat{A}^{+},\hat{C}^{-}]=0,\qquad &&
 [\hat{A}^{+},\hat{C}]=\frac{1}{2} \hat{A}^{+},& \\
&[\hat{A}^{-},\hat{B}^{+}]=\hat{C}^{+}, \qquad &&
 [\hat{A}^{-},\hat{B}^{-}]=0, \qquad &&
 [\hat{A}^{-},\hat{B}]=\frac{1}{2} \hat{A}^{-}, &\\
&[\hat{A}^{-},\hat{C}^{+}]=0, \qquad&&
 [\hat{A}^{-},\hat{C}^{-}]=-\hat{B}^{-}, \qquad &&
 [\hat{A}^{-},\hat{C}]=-\frac{1}{2} \hat{A}^{-}, & \\
&[\hat{A}, \hat{B}^{+}]=\frac{1}{2} \hat{B}^{+}, \qquad&&
 [\hat{A},\hat{B}^{-}]=-\frac{1}{2} \hat{B}^{-}, \qquad&&
 [\hat{A}, \hat{B}]=0, & \\
&[\hat{A}, \hat{C}^{+}]=-\frac{1}{2} \hat{C}^{+}, \qquad&&
 [\hat{A}, \hat{C}^{-}]=\frac{1}{2} \hat{C}^{-}, \qquad&&
 [\hat{A}, \hat{C}]=0, &  \\
&[ \hat{B}^{+},\hat{C}^{+}]=0, \qquad &&
 [\hat{B}^{+},\hat{C}^{-}]=-\hat{A}^{+}, \qquad&&
 [\hat{B}^{+},\hat{C}]=-\frac{1}{2} \hat{B}^{+}, & \\
&[\hat{B}^{-},\hat{C}^{+}]=\hat{A}^{-}, \qquad&&
 [ \hat{B}^{-},\hat{C}^{-}]=0, \qquad &&
 [\hat{B}^{-},\hat{C}]=\frac{1}{2} \hat{B}^{-}, & \\
&[\hat{B},\hat{C}^{+}]=\frac{1}{2} \hat{C}^{+}, \qquad &&
 [\hat{B}, \hat{C}^{-}]=-\frac{1}{2} \hat{C}^{-}, \qquad&&
[\hat{B}, \hat{C}]=0 .&
\end{alignat*}
The
second order Casimir operator of $su(2,1)$  is
\begin{gather*}
{\cal C} = \hat A^+\hat A^- - \hat B^+\hat B^- - \hat C^+\hat C^- +
\frac23\big(
\hat A^2+  \hat B^2 +   \hat C^2\big) -
(\hat A+\hat B+\hat C).
\end{gather*}
Note  that in our dif\/ferential realization
\[
\hat A - \hat B + \hat C =0,
\]
and that there is another generator,
\begin{gather*}
{\cal C}'= l_1+l_2-l_0,
\end{gather*}
commuting with the rest of generators of $su(2,1)$. Hence
\[
\langle \hat{A}^{\pm},\hat{A},\hat{B}^{\pm},
\hat{B},\hat{C}^{\pm},\hat{C}\rangle \oplus \langle {\cal C}' \rangle
 \approx   u(2,1) .
 \]
 The Hamiltonian \eqref{hamiltoniansu(2,1)} can be rewritten in terms of ${\cal C}$ and
 ${\cal C}'$ as
 \begin{gather*}
H_{\ell}  = -4  {\cal C}+ \frac13   {{\cal C}'}^2 - \frac{15}{4}\\
\phantom{H_{\ell}}{} = -4\left( \hat A^+\hat A^- - \hat B^+\hat B^- - \hat C^+\hat C^- +
\frac23\big(
\hat A^2+  \hat B^2 +   \hat C^2\big) -
(\hat A+\hat B+\hat C) \right)+ \frac13   {{\cal C}'}^2 - \frac{15}{4}.
\end{gather*}
The quadratic operators $ \hat A^+\hat A^-$, $ \hat B^+\hat B^-$ and
$ \hat C^+\hat C^-$ commute with the Hamiltonian and they are constants of motion. However, they do not commute among
themselves giving cubic expressions
\[\
[ \hat A^+\hat A^-,  \hat B^+\hat B^-]= -
[ \hat A^+\hat A^-,  \hat C^+\hat C^-]= -
[  \hat B^+\hat B^-, \hat C^+\hat C^-]= \hat A^+\hat C^+\hat B^- -
\hat B^+\hat C^-\hat A^-.
\]
However, as we will see later, they generate a quadratic algebra.

The eigenfunctions of the Hamiltonians $H_{\ell}$,
that have the same energy, support an IUR
of~$su(2,1)$ characterized by a value of
${\cal C}$ and other of  ${\cal C}'$.
These representations can be obtained, as usual, starting from
a fundamental state simultaneously annihilated by the lowering opera\-tors~$\hat A^-$, $\hat C^-$ and  $\hat B^-$
\[
 \hat A^-_{\ell}\Phi^0_{\ell} = \hat C^-_{\ell} \Phi^0_{\ell} =
\hat
B^-_{\ell} \Phi^0_{\ell} =0  .
\]
Solving these equations  we f\/ind
\begin{gather}\label{eigenstatesu21}
\Phi^0_{\ell}(\xi,\theta) =
N (\cos \theta)^{l_0+1/2}
(\sin \theta)^{1/2} (\cosh \xi)^{l_2+1/2}(\sinh \xi)^{l_0+1},
\end{gather}
where $\ell= (l_0, 0, l_2)$ and $N$ is a normalization constant.
From previous inequalities  the parame\-ters~$l_0$,~$l_2$
of $\Phi^0_{\ell}$ must satisfy
\[
(l_0+l_2)<-3/2 ,\qquad l_0\geq -1/2 .
\]
In order to guarantee the normalization of
$\Phi^0_{\ell}$ using the invariant measure
the values of the parameters $l_0$ and $l_2$ have to verify
\[
(l_0+l_2)<-5/2 .
\]
Note that the  state $\Phi^0_{\ell}$ supports also
IUR's of the subalgebras $su(2)$ (generated by $\hat A^{\pm}$ with
the weight $j=l_0/2$)  and $su(1,1)$ (generated by $\hat C^{\pm}$
with $j'_2=-l_2/2$).

The energies of the fundamental states $\Phi^0_{\ell}(\xi,\theta)$ are
obtained from $H_{\ell}$
taking into account
the expressions for the Casimir operators ${\cal C}$ and
${\cal C}'$
\[
H_{\ell} \Phi^0_{\ell} = - (l_0+l_2+3/2)(l_0+l_2+5/2) \Phi^0_{\ell}
\equiv E^0_{\ell} \Phi^0_{\ell}   .
\]
From $\Phi^0_{\ell}$ we can get the  other eigenfunctions in the
$su(2,1)$ representation using the raising operators
$\hat A^+$, $\hat B^+$, $\hat C^+$, all of them sharing the same energy eigenvalue
$E^0_{\ell}$.
Notice that  the expression  for  $E^0_{\ell}$  depends on
$l_0+l_2$, hence the states in the family of IUR's derived from fundamental states $\Phi^0_{\ell}(\xi,\theta)$,  sharing the same value of $l_0+l_2$, also shall have the same energy eigenvalue.
The energy $E^0_{\ell}$ corresponding to bound states is negative
and the set of such bound states for each Hamiltonian $H_{\ell}$ is f\/inite.

\begin{figure}[t]
\centerline{\includegraphics[scale=0.5]{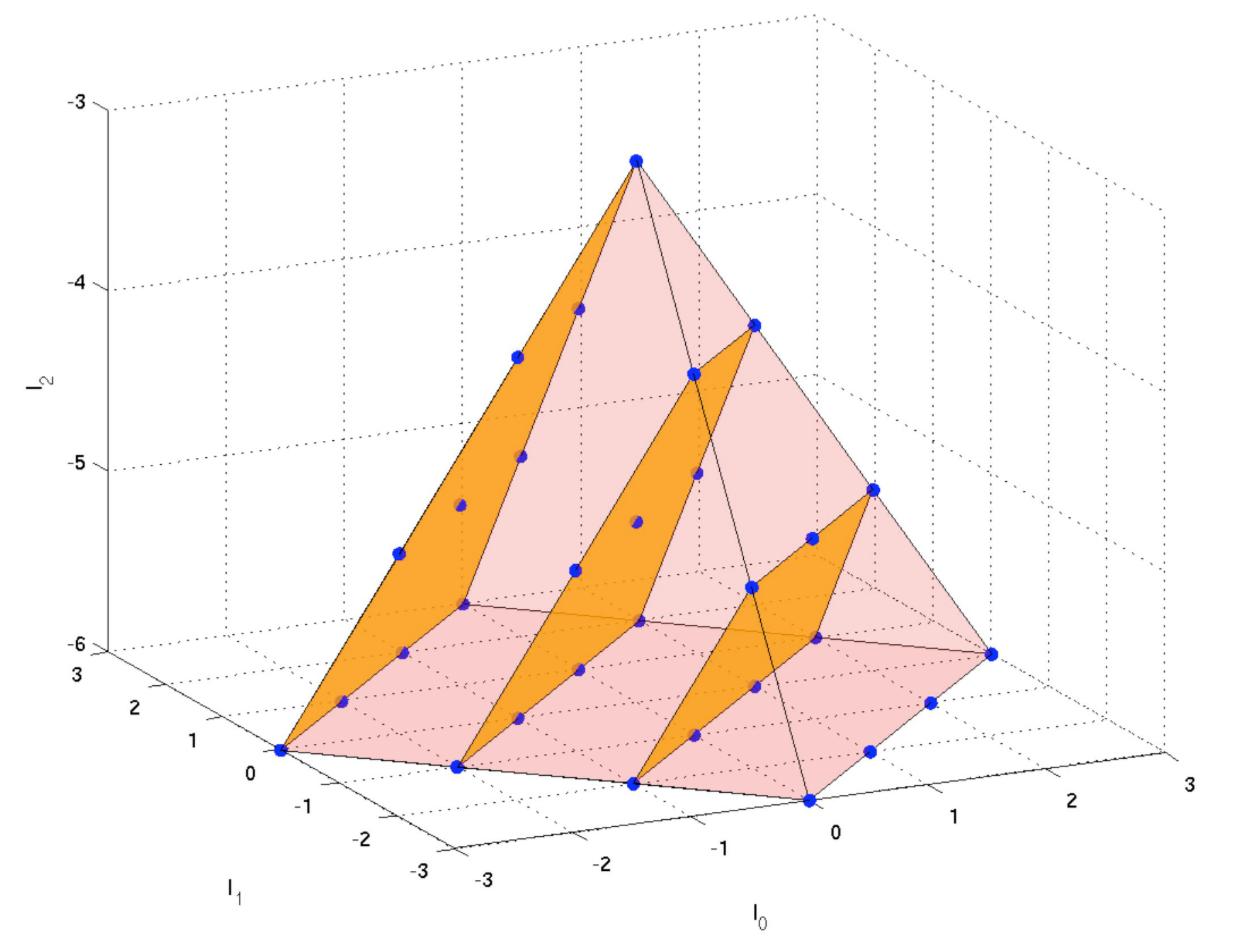}}

\caption{States of IUR's of $su(2,1)$ sharing
the same energy and
represented by points in the three dark planes associated
to $\Phi^0_{\ell}$ with $\ell=(0,0,-3)$,
$\ell=(1,0,-4)$ and $\ell=(2,0,-5)$.} \label{fig1}
\end{figure}

In Fig.~\ref{fig1}
we display the states of some IUR's of $su(2,1)$ by points
$(l_0,l_1,l_2)\in \mathbb{R}^3$ linked to the ground state
$\Phi^0_{\ell}$, characterized by   $(l_0,0,l_2)$, by
the raising operators  $\hat A^+$ and $\hat C^+$.
The points associated to a IUR
are in a 2D plane (f\/ixed by the particular value $l_0+l_2=-3$ of
${\cal C}'$) and, obviously,  the other IUR's are described by points in
parallel 2D  planes. These parallel planes are placed inside a
tetrahedral unbounded pyramid whose basis extends towards
$ -\infty$ along the axis $l_2$.

On the other hand, there exist some points (in the parameter space of parameters ($l_0, l_1,l_2$)) which
are degenerated because  they correspond to an eigenspace with
dimension bigger than~$1$.  For instance, let us consider
the  representation
characterized by the fundamental state $\Phi^0_{\ell}$ where
 $\ell=(0,0,-3)$: its points
 lie in a triangle and  are nondegenerated.
The IUR corresponding to the ground
state with $\ell'=(1,0,-4)$ has eigenstates with the same
energy, \mbox{$E= -(-3+3/2)(-3+5/2)$}, as the previous one,  since
both share the same value of $l_0+l_2=-3$.
The eigenstates corresponding to $\ell''=(0,0,-5)$, inside this representation, may be obtained in two ways:
\[
\Phi^2_{(0,0,-5)} = \hat C^+ \hat A^+ \Phi^0_{(0,0,-3)},\qquad
\tilde\Phi^2_{(0,0,-5)} = \hat A^+ \hat C^+ \Phi^0_{(0,0,-3)}   .
\]
We have two   independent states  spanning
a 2-dimensional  eigenspace of
the  Hamiltonian $H_{(0,0,-5)}\!$ for that eigenvalue of the energy
$E= -(-3+3/2)(-3+5/2)$.
The ground state for  $H_{(0,0,-5)}$ is given by the wavefunction $\Phi^0_{(0,0,-5)}$ and its
energy is $E^0_{(0,0,-5)}= - (-5+3/2)(-5+5/2)$.

In a similar way it is possible to  obtain the degeneration of
higher excited levels in the discrete spectrum of the Hamiltonians. Thus,
the $n$-excited level, when it exists,
has associated an $n$-dimensional eigenspace.

\subsubsection[The complete symmetry algebra ${so(4,2)}$]{The complete symmetry algebra $\boldsymbol{so(4,2)}$}
\label{completeso42}

By simple inspection one can see that the  Hamiltonian
$H_{\ell}$ \eqref{hamiltoniansu(2,1)} is invariant under ref\/lections
in the space of parameters $(l_0,l_1,l_2)$
\begin{gather}
I_0: \ (l_0,l_1,l_2)\to (-l_0,l_1,l_2),\qquad
I_1: \ (l_0,l_1,l_2)\to (l_0,-l_1,l_2),\nonumber\\
I_2: \ (l_0,l_1,l_2)\to (l_0,l_1,-l_2).\label{reflections}
\end{gather}
These operators  generate by conjugation other sets of intertwining operators from the ones already def\/ined. Thus,
\[
I_0: \ \{ \hat A^{\pm},\hat A \} \longrightarrow
\{ \tilde A^{\pm}= I_0 \hat A^{\pm}I_0, \tilde A = I_0 \hat A I_0 \}  ,
\]
where
\[
\tilde A_{l_{0},l_{1}}^{\pm}=\pm\partial_{\theta}-(-l_{0}+1/2)
 \tan{\theta}
+(l_{1}+1/2) \cot{\theta} ,\qquad
\tilde \lambda_{l_{0},l_{1}}=(1-l_{0}+l_{1})^2.
\]
They act on the eigenfunctions of the Hamiltonians
 \eqref{hamiltoniansu(2,1)a} in the following way
\[
\tilde A_{l_{0},l_{1}}^{-}: \ f_{l_{0},l_{1}}\rightarrow
f_{l_{0}-1,l_{1}+1},\qquad
\tilde A_{l_{0},l_{1}}^{+}: \ f_{l_{0-1},l_{1}+1}\rightarrow
f_{l_{0},l_{1}}.
\]
In these conditions, we can def\/ine global operators $\tilde A^\pm$ as we made before. Then,   $\tilde A^\pm$ together with
$
\tilde{A} f_{l_{0},l_{1}}:=-\frac{1}{2} (-l_{0}+l_{1}) f_{l_{0},l_{1}}
$
close a second $\widetilde{su}(2)$.

In a similar way  new sets of operators
$\{\tilde B^{\pm},\tilde B\}$ and
$\{\tilde C^{\pm},\tilde C\}$ closing $\widetilde{su}(1,1)$  algebras, can also be def\/ined
\begin{gather*}
I_0: \ \{A^\pm, A;   B^\pm, B;   C^\pm, C \} \longrightarrow
\{\tilde A^\pm, \tilde A;  \tilde B^\pm, \tilde B;   C^\pm, C \},
\\
I_1: \ \{A^\pm, A;   B^\pm, B;   C^\pm, C \} \longrightarrow
\{\tilde A^\mp, -\tilde A;   B^\pm,  B;  \tilde C^\pm, \tilde C \},
\\
I_2: \ \{A^\pm, A;  B^\pm, B;   C^\pm, C \} \longrightarrow
\{  A^\pm,   A;  \tilde B^\mp,-\tilde B;  -\tilde C^\mp, -\tilde C \}.
\end{gather*}

The whole set of the operators
$\{A^\pm,\tilde A^\pm,
B^\pm, \tilde B^\pm, C^\pm,\tilde C^\pm\}$
together with  the set of diagonal operators $\{L_0, L_1, L_2\}$,  def\/ined by
\[
L_i \Psi_\ell = l_i \Psi_\ell,
\]
span a Lie algebra of rank three: $o(4,2)$.
The Lie commutators of $o(4,2)$ can be easily derived from those of $su(2,1)$   and the action of the ref\/lections.
It is obvious, by construction, that all these generators link eigenstates of  Hamiltonians $H_\ell$ with the same eigenvalue.

The fundamental state $\Psi^0_\ell$ for
$so(4,2)$ will be annihilated by all  the lowering operators
 \[
A^-_{\ell}\Psi^0_{\ell} =\tilde A^-_{\ell}\Psi^0_{\ell}=
C^-_{\ell} \Psi^0_{\ell} =\tilde C^-_{\ell} \Psi^0_{\ell}=
B^-_{\ell} \Psi^0_{\ell} = \tilde B^-_{\ell} \Psi^0_{\ell} = 0  .
\]
This state should be a particular case of the state given by expression \eqref{eigenstatesu21}, i.e.\ it should be
also invariant under the $l_0$-ref\/lection,
\[
\Phi^0_{(l_0=0,l_1=0,l_2)}(\xi,\theta) =
N (\cos \theta)^{1/2}
(\sin \theta)^{1/2} (\cosh \xi)^{l_2+1/2}\sinh \xi,
\]
where $l_2<-5/2$.
This point $(l_0=0,l_1=0,l_2)$ in the parameter space, for the cases displayed in  Fig.~\ref{fig1}, corresponds to the top vertex of the pyramid,
from which all the other points plotted  can
be obtained with the help of raising operators.
Such points correspond to an IUR of  $so(4,2)$
algebra that includes the series of IUR's of  $su(2,1)$.

\begin{figure}[t]
\centerline{\includegraphics[scale=0.5]{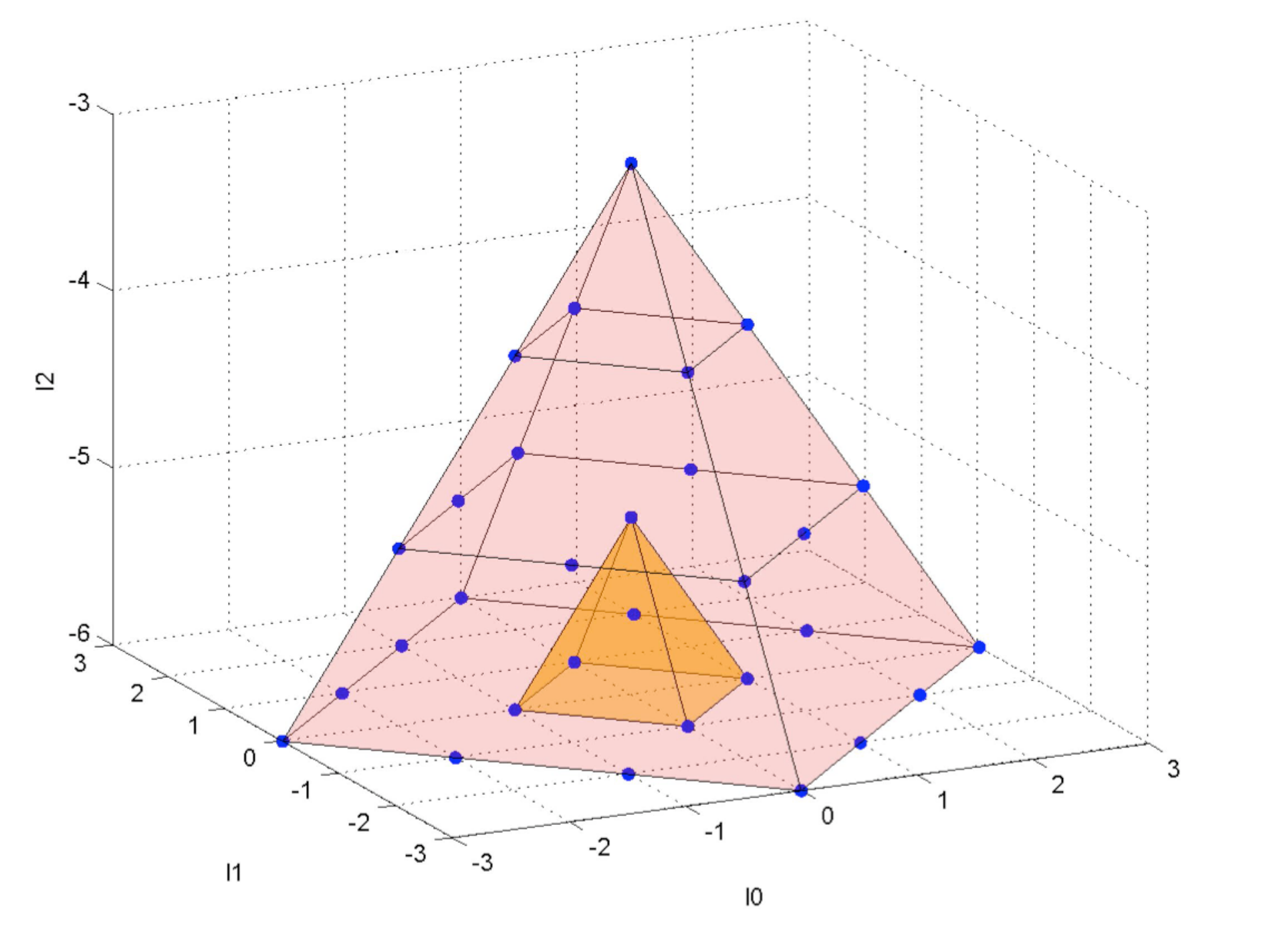}}

\caption{Two pyramids associated to the
same IUR of $so(4,2)$. The points of the faces
of the exterior pyramid (with vertex $(0,0,-3)$)
represent non-degenerated levels. The  exterior faces of the
inner pyramid (vertex $(0,0,-5)$)
are f\/irst excited
double-degenerated levels.} \label{fig2}
\end{figure}

Fixed the IUR of $so(4,2)$ corresponding to a value of $\ell=(0,0,l_2)$ such that
$-7/2\leq l_2 <-5/2$, then the points on the surface of the
associated pyramid in the parameter space correspond to
non-degenerated ground levels of their respective Hamiltonians.
This `top' pyramid includes inside other `lower' pyramids with vertexes
at the points $\ell_n = (0,0,l_2-2n)$.
Each point
on the surface of an inner pyramid associated to $\ell_n$ represents
an $n$-excited level $n$-fold degenerated of the IUR associated to
$\ell$ (see Fig.~\ref{fig2}).

\subsection[Superintegrable quantum   $u(3)$-system]{Superintegrable quantum   $\boldsymbol{u(3)}$-system}
\label{su3sqsystems}

In this case we consider the quantum Hamiltonian
\begin{gather}\label{hamiltoniansu(3)}
H_{\ell} = - \left(J_0^2 +
        J_1^2 + J_2^2 \right)
      +\frac{l_0^2-1/4}{(s_0)^2} + \frac{l_1^2-1/4}{(s_1)^2}+
        \frac{l_2^2-1/4}{(s_2)^2},
\end{gather}
where $\ell=(l_0,l_1,l_2)\in \R^3$, $J_i =-\epsilon_{ijk}s_j\p_k$
($i=0,1,2$) and its conf\/iguration space is the 2-sphere
\[
{\cal S}^2\equiv (s_0)^2+  (s_1)^2 +  (s_2)^2 = 1,\qquad
(s_0,s_1,s_2)\in \R^3.
\]
In spherical coordinates
\begin{gather}\label{sphericalcoordiantes1}
   s_0 = \cos\phi \cos\theta,\!\qquad
   s_1 = \cos\phi  \sin\theta,\!\qquad
   s_2  = \sin\phi ,\!\qquad \phi\in[-\pi/2,\pi/2],\quad \theta\in [0,2\pi] ,\!\!\!
\end{gather}
that parametrize ${\cal S}^2$, the eigenvalue problem
$H_\ell \Psi = E  \Psi$
takes the expression
\[
\left[ -\partial_{\phi}^2 +
    \tan\phi  \partial_{\phi}  +
    \frac{l_2^2-1/4}{\sin^{2}(\phi)}
 + \frac{1}{\cos^{2}\phi}
\left[ - \partial_{\theta}^2 +
    \frac{l_0^2-1/4}{\cos^{2}\theta} +
    \frac{l_1^2-1/4}{\sin^{2}\theta}\right]
\right]\Psi = E  \Psi .
\]
Taking  solutions separated in the variables $\theta$ and $\phi$ as
${\bf  \Psi}(\theta,\phi) =f(\theta)  g(\phi) $ we f\/ind
 \begin{gather}\label{sisn21}
H^\theta_{l_0,l_1}f(\theta)\equiv
\left[ -
\partial_{\theta}^2 +
    \frac{l_0^2-1/4}{\cos^{2}\theta} +
    \frac{l_1^2-1/4}{\sin^{2}\theta}\right] f(\theta) =
\alpha f(\theta) ,
\\
 \left[ -\partial_{\phi}^2 +
    \tan\phi  \partial_{\phi} + \frac{\alpha}{\cos^{2}\phi}+
    \frac{l_2^2-1/4}{\sin^{2}\phi} \right] g(\phi) =
    E g(\phi)   ,\nonumber
\end{gather}
with $\a >0$ a separating constant. Note that equation \eqref{sisn21} is equal to equation \eqref{hamiltoniansu(2,1)a} corresponding to  the
$su(2,1)$ case.

Following the procedure of the previous case of $so(2,1)$ we can factorize the Hamiltonian \eqref{sisn21} in terms of operators $A_n^\pm$ like those of expression \eqref{operadoresA},
 obtaining, f\/inally, a $su(2)$ algebra.

We can f\/ind other two sets of spherical coordinates, that parametrize the sphere ${\cal S}^2$ and that separate the Hamiltonian \eqref{sisn21}, Thus, we get  two new sets of intertwining opera\-tors~$B_n^\pm$ and~$C_n^\pm$,   like in the $su(2,1)$ case. In this way, we can construct an algebra $u(3)$  and using ref\/lection operators, acting in the space of the parameters of the Hamiltonian \eqref{hamiltoniansu(3)}, this algebra is  enlarged to $so(6)$.

These three sets of operators are related, as we saw in Section~\ref{su21sqsystems},
 with  three sets of (spheri\-cal) coordinates that we can take in the 2-sphere immersed in a 3-dimensional ambient space with cartesian axes
$\{s_0,s_1,s_2\}$.   Since the coordinates $(s_0,s_1,s_2)$ play a symmetric role in the
Hamiltonian  \eqref{hamiltoniansu(3)}, we will take their cyclic rotations to get two other intertwining sets.
Thus, we take the spherical
coordinates choosing as `third axis'
$s_1$ instead of $s_2$ as in
\eqref{sphericalcoordiantes1},
\begin{gather*}
   s_2 = \cos\psi \cos\xi, \qquad
   s_0 = \cos\psi  \sin\xi, \qquad
   s_1  = \sin\psi,   \qquad \psi\in[-\pi/2,\pi/2],\quad \xi\in [0,2\pi] .
\end{gather*}
The corresponding  intertwining operators
$B_{(l_0,l_1,l_2)}^{\pm}$ are def\/ined in a similar  way to
$A_{(l_0,l_1,l_2)}^{\pm}$. The explicit
 expressions for the new set in terms of the initial coordinates
($\theta, \phi$) (\ref{sphericalcoordiantes1}) are
\[\label{bls}
 B_{(l_0,l_1,l_2)}^{\pm}= \pm (\sin\theta
 \tan\phi \partial_\theta +\cos\theta \partial_\phi  ) -  (l_2{+}1/2)
 \cos\theta \cot \phi + (l_0{+}1/2) \sec  \theta \tan\phi .
 \]
 The spherical
coordinates around the $s_0$ axis are
\begin{gather*}
   s_1 = \cos \beta  \cos \eta ,\qquad
   s_2 = \cos \beta   \sin \eta ,\qquad
   s_0  = \sin\beta.
\end{gather*}
We obtain a new pair of operators,
that written in terms of the original  variables $(\theta,\phi)$ are
\[
C_{(l_0,l_1,l_2)}^{\pm}= \pm (\cos\theta \tan \phi \partial_{\theta}
-\sin\theta \partial_{\phi}) +  (l_1{-}1/2) \,{\rm cosec}\,\theta
 \tan \phi + (l_2{+}1/2) \sin\theta \cot\phi .
\]
They  intertwine  the Hamiltonians in the following way
 \[
 C^{-}_{(l_0,l_1,l_2)}H_{(l_0,l_1,l_2)}=
H_{(l_0,l_1-1,l_2+1)}C^{-}_{(l_0,l_1,l_2)},\qquad
C^{+}_{(l_0,l_1,l_2)}H_{(l_0,l_1-1,l_2+1)}=H_{(l_0,l_1,l_2)}C^{+}_{(l_0,l_1,l_2)} .
\]
The free-index or `global' operators  close a third $su(2)$.

All these transformations  $\{ A^{\pm}, A, B^{\pm}, B, C^{\pm}, C \}$ (where $A-B+C=0$) span  an  algebra $su(3)$, whose  Lie commutators are
\begin{alignat*}{4}
&[\hat{A}^{+}, \hat{A}^{-}]=2 A, \qquad &&
 [\hat{A},\hat{A}^{\pm}]=\pm\hat{A}^{\pm}, \qquad&&
 [\hat{B}^{+}, \hat{B}^{-}]=2 B,& \\
&[\hat{B},\hat{B}^{\pm}]=\pm\hat{B}^{\pm}, \qquad&&
 [\hat{C},\hat{C}^{\pm}]=\pm\hat{C}^{\pm}, \qquad&&
 [\hat{C}^{+}, \hat{C}^{-}]=2 C, & \\
&[\hat{A}^{+}, \hat{B}^{+}]=0, \qquad &&
 [\hat{A}^{+},\hat{B}^{-}]=\hat{C}^{-}, \qquad&&
 [\hat{A}^{+},\hat{B}]=-\frac{1}{2} \hat{A}^{+}, & \\
&[\hat{A}^{+},\hat{C}^{+}]=-\hat{B}^{+},\qquad &&
 [\hat{A}^{+},\hat{C}^{-}]=0,\qquad &&
 [\hat{A}^{+},\hat{C}]=\frac{1}{2} \hat{A}^{+},& \\
&[\hat{A}^{-},\hat{B}^{+}]=-\hat{C}^{+},\qquad &&
 [\hat{A}^{-},\hat{B}^{-}]=0,\qquad&&
 [\hat{A}^{-},\hat{B}]=\frac{1}{2} \hat{A}^{-},&\\
&[\hat{A}^{-},\hat{C}^{+}]=0,\qquad &&
 [\hat{A}^{-},\hat{C}^{-}]=\hat{B}^{-},\qquad &&
 [\hat{A}^{-},\hat{C}]=-\frac{1}{2} \hat{A}^{-}, & \\
&[\hat{A}, \hat{B}^{+}]=\frac{1}{2} \hat{B}^{+}, \qquad&&
 [\hat{A},\hat{B}^{-}]=-\frac{1}{2} \hat{B}^{-},\qquad&&
 [\hat{A}, \hat{B}]=0, & \\
&[\hat{A}, \hat{C}^{+}]=-\frac{1}{2} \hat{C}^{+}, \qquad&&
 [\hat{A}, \hat{C}^{-}]=\frac{1}{2} \hat{C}^{-},\qquad &&
 [\hat{A}, \hat{C}]=0, & \\
&[ \hat{B}^{+},\hat{C}^{+}]=0, \qquad&&
 [\hat{B}^{+},\hat{C}^{-}]=-\hat{A}^{+}, \qquad&&
 [\hat{B}^{+},\hat{C}]=-\frac{1}{2} \hat{B}^{+},& \\
&[\hat{B}^{-},\hat{C}^{+}]=\hat{A}^{-}, \qquad&&
 [ \hat{B}^{-},\hat{C}^{-}]=0, \qquad&&
 [\hat{B}^{-},\hat{C}]=\frac{1}{2} \hat{B}^{-}, &\\
&[\hat{B},\hat{C}^{+}]=\frac{1}{2} \hat{C}^{+}, \qquad&&
 [\hat{B}, \hat{C}^{-}]=-\frac{1}{2} \hat{C}^{-}, \qquad&&
 [\hat{B}, \hat{C}]=0 .&
\end{alignat*}
The second order  Casimir operator of $su(3)$  is given by
\begin{gather}\label{cassu3}
{\cal C}= A^+A^- + B^+B^- + C^+C^- + \frac23 A(A-3/2)
+ \frac23 B(B-3/2) + \frac23 C(C-3/2) .
 \end{gather}
We obtain  a $u(3)$ algebra by adding the central diagonal operator \begin{gather}\label{casu3}
D:= l_0-l_1-l_2 .
\end{gather}
The global
operator convention can be adopted for the Hamiltonians  in the
$u(3)$-hierarchy by
def\/ining its action on the eigenfunctions $\Phi_{(l_1,l_2,l_3)}$ of
$H_{(l_1,l_2,l_3)}$ by
$H\Phi_{(l_1,l_2,l_3)}:= H_{(l_1,l_2,l_3)}
\Phi_{(l_1,l_2,l_3)}$.
Then,  $H$   can be expressed in terms of both Casimir operators,
\eqref{cassu3} and \eqref{casu3}, as
\begin{gather}\label{esp}
H= 4 {\cal C}-\frac13  D^2+\frac{15}4 .
 \end{gather}
  Hence, the  Hamiltonian can be written as a certain quadratic function
of the operators~$A^\pm$,~$B^\pm$ and $C^\pm$ generalizing the usual
factorization for one-dimensional systems plus a constant since   in the representation that we are using the operators $A$, $B$, $C$ are diagonal depending on the parameters $l_0$, $l_1$, $l_2$,
\begin{gather*}
H= 4 (A^+A^- + B^+B^- + C^+C^-)+\text{cnt} .
 \end{gather*}
The quadratic operators $A^+A^-$, $B^+B^-$, $C^+C^-$ commute with $H$ but do not commute among themselves
\[
[A^+A^- , B^+B^- ]=-[ A^+A^- ,C^+C^-]=[B^+B^- , C^+C^-]
=-A^+C^+B^-+B^+ C^-A^-.
\]

The intertwining operators can help also in supplying the
elementary integrals of motion. We have two kinds of integrals:
(i) second order constants, def\/ined by the quadratic operators
$X_1= A^+A$, $X_2 = B^+B$, $X_3= C^+C$; and (ii) third order constants
def\/ined by cubic operators: $Y_1=  A^+C^+B^-$, $Y_2 =B^+C^-A^-= (Y_1)^+$.
Of course, since this system is superintegrable, there are only three functionally independent constants of motion, for instance
$X_1$, $X_2$, $X_3$. This set of symmetries $\{X_i,Y_j\}$ closes a quadratic
algebra, as it is well known from many references
\cite{daskaloyannis01,kuru02, zhedanov_92,bambah_07,kalnins96,kalnins97}. The commutators in this
case are
\begin{gather*}
[X_1 , X_2 ]=-[ X_1 ,X_3]=[X_2 , X_3]
=-Y_1 + Y_2,\\
 [X_1,Y_1]= X_1 X_2 - X_1 X_3-2(A-1)Y_1,\\
 [X_1,Y_2] = - X_2 X_1 + X_3 X_1+ 2 (A - 1) Y_2,  \\
 [X_2,Y_1] =  X_1 X_2- X_2 X_3 - (1 + 2 B) Y_1+ Y_2 -2 C X_2, \\
 [X_2,Y_2] = - X_2 X_1 + X_3 X_2 + (1 + 2 B) Y_2- Y_1+2 C X_2,
  \\
 [X_3,Y_1]= - X_1 X_3 - X_2 X_3+ 2 C Y_1 -2 C X_2  + Y_2,\\
 [X_3,Y_2]=  X_3 X_1 + X_3 X_2 - 2 C Y_2 + 2 C X_2  - Y_1, \\
  [Y_1,Y_2]=2 ( -C X_1 X_2 + B X_1 X_3 + A X_2 X_3 +
   (B+C) Y_1 - A Y_2 + 2 AC X_2).
   \end{gather*}
Remark that the operators $\{A,B,C\}$ are diagonal with f\/ixed values
for each Hamiltonian.

One can show that the eigenstates of this Hamiltonian hierarchy
are connected to the IUR's of $u(3)$. Fundamental states $\Phi$
annihilated by $A^-$ and
$C^-$ (simple roots of $su(3)$),
\[
A_\ell^-\Phi_\ell
= C_\ell^-\Phi_\ell =0 ,
\]
only  exist  when $l_1=0$. Their explicit form is
\[\label{ground}
\Phi_\ell(\theta,\phi)= N
 \cos^{l_0{+}1/2}\theta \sin^{1/2}\theta \cos^{l_0+1}\phi \sin^{l_2{+}1/2}\phi ,
\]
whit $N$  a normalizing constant. The diagonal operators act
on them  as
\begin{gather}
A  \Phi_\ell = -l_0/2  \Phi_\ell,\qquad   l_0 = m,\quad l_1=0,\quad  m=0,1,2,\dots,\nonumber\\
C  \Phi_\ell = -l_2/2  \Phi_\ell,\qquad   l_2 = n,\quad   n=0,1,2,\dots.\label{mn}
\end{gather}
This shows that $\Phi_\ell$ is the lowest state of the
IUR $j_1=m/2$ of the  subalgebra $su(2)$ generated by $\{A^\pm,A\}$,
and of the IUR $j_2=n/2$ of the subalgebra $su(2)$ spanned by
$\{C^\pm,C\}$. Such a~$su(3)$-representation  will be denoted
$(m,n)$, $m,n\in\Z^{\geq 0}$. The points (labelling the states) of this
representation obtained from $\Phi_\ell$   lie on the plane
$D=m-n$ inside the $\ell$-parameter space.

The energy for the states of the IUR, determined by the lowest
state (\ref{mn}) with
  parameters $(l_0,0,l_2)$,
is given (\ref{esp}) by
\begin{gather*}
E=(l_0 + l_2 + 3/2) ( l_0  + l_2 + 5/2)=(m
+ n + 3/2) ( m  + n + 5/2) .
\end{gather*}
Note that the IUR's labelled by $(m,n)$
with the same value $m+n$ are associated to states with the same energy (iso-energy representations).
This degeneration will be broken  using  $so(6)$.

Making use of some relevant discrete symmetries, following the procedure of
Section~\ref{completeso42},
the dynamical algebra $u(3)$ can be enlarged to $so(6)$.

The
Hamiltonian $H_{(l_0,l_1,l_2)}$  (\ref{hamiltoniansu(3)}) is invariant under ref\/lections  \eqref{reflections}  in
the parameter space $(l_0,l_1,l_2)$.
These symmetries, $I_i$, can be directly implemented in
the eigenfunction space, gi\-ving by  conjugation
another set
of intertwining operators $\{\tilde X = I_i   X   I_i,\;   i=0,1,2\}$
  closing an isomorphic Lie algebra
$\tilde u(3)$.
They are
\begin{gather*}
\{A^\pm,B^\pm,C^\pm\} \stackrel{I_0} \longrightarrow   \{\tilde
A^\mp,\tilde B^\mp,C^\pm\},\\
\{A^\pm,B^\pm,C^\pm\} \stackrel{I_1} \longrightarrow   \{\tilde
A^\pm, B^\pm,\tilde C^\pm\},\\
\{A^\pm,B^\pm,C^\pm\} \stackrel{I_2} \longrightarrow
\{A^\pm,\tilde B^\pm, \tilde C^\mp\} .
\end{gather*}
The set $\{A^\pm,\tilde A^\pm,
B^\pm, \tilde B^\pm, C^\pm,\tilde C^\pm, A,\tilde A ,B,
\tilde B,C,\tilde C\}$ closes a Lie algebra of rank~3: $so(6)$. However, instead of the six non-independent  generators $A,\tilde A, B,\tilde B, C,\tilde C$   it is enough to consider three independent
diagonal operators $L_0$, $L_1$, $L_2$ def\/ined by
$ L_i
\Psi_{(l_0,l_1,l_2)} \equiv l_i  \Psi_{(l_0,l_1,l_2)}$.
 The Hamiltonian can be expressed in terms of the $so(6)$-Casimir operator
  by means of the `symmetrization' of the $u(3)$-Hamiltonian (\ref{esp})
  \begin{gather*}
H =\{A^+,A^-\}+\{B^+,B^-\}+\{C^+,C^-\}
+ \{\tilde A^+, \tilde A^-\}+\{\tilde B^+, \tilde B^-\}+ \{\tilde
C^+, \tilde C^-\}\\
\phantom{H=}{} + {L_0}^2+{L_1}^2+{L_2}^2 +\frac{41}{12}.
\end{gather*}

The intertwining generators of $so(6)$ give rise to larger
3-dimensional Hamiltonian hierarchies
\begin{gather}\label{so6hierarchies}
\{H_{(l_0+m+p,l_1+m-n-p,l_2+n)} \},\qquad m,n,p\in\Z ,
\end{gather}
each one including a class of the previous ones coming from $u(3)$.

The eigenstates of these $so(6)$-hierarchies can be classif\/ied
in terms of $so(6)$ representations whose fundamental states
$\Phi^0_\ell$ are determined  by
\[
A^-\Phi^0_\ell=\tilde A^- \Phi^0_\ell=C^- \Phi^0_\ell,
\]
and whose explicit expressions are
\[
\Phi^0_\ell(\theta,\phi)= N \cos^{1/2}\theta  \sin^{1/2}\theta
\cos \phi \sin^{l_2+1/2}\phi .
\]
They are characterized by the eigenvalues of the diagonal operators $L_i$
\[
L_0 \Phi^0_\ell=L_1 \Phi^0_\ell=0,\qquad
L_2 \Phi^0_\ell=n  \Phi^0_\ell ,\qquad  n\in\Z ^+.
\]
We obtain two classes of symmetric IUR,s of $so(6)$ that according to \eqref{so6hierarchies} we summarize as
 \begin{alignat*}{5}
& \text{even IURs} : \ \  && l_0=0,\  l_1=0,\  l_2=0,\qquad &&
  \{ H_{(m+p,m-n-p,n)} \},\qquad && q m,n,p\in\Z ,& \\
& \text{odd IURs} : \ \   &&  l_0=0,\  l_1=0,\  l_2=1,\qquad &&
  \{ H_{(m+p,m-n-p,1+n)} \},\qquad  && m,n,p\in\Z  .&
\end{alignat*}

\begin{figure}[t]
\centerline{\includegraphics[scale=0.67]{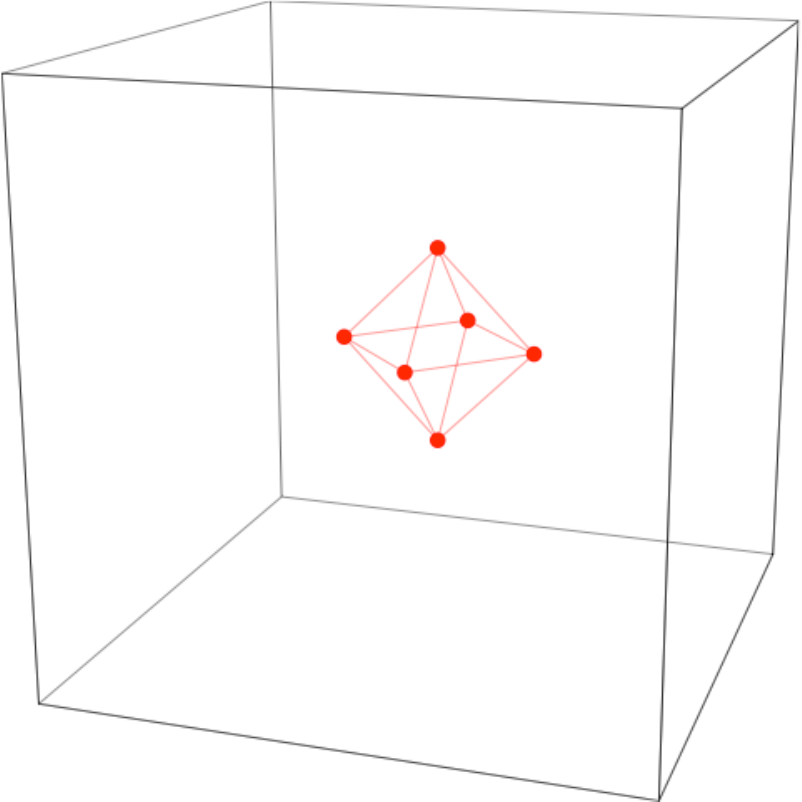}\hskip1.5cm
\includegraphics[scale=0.67]{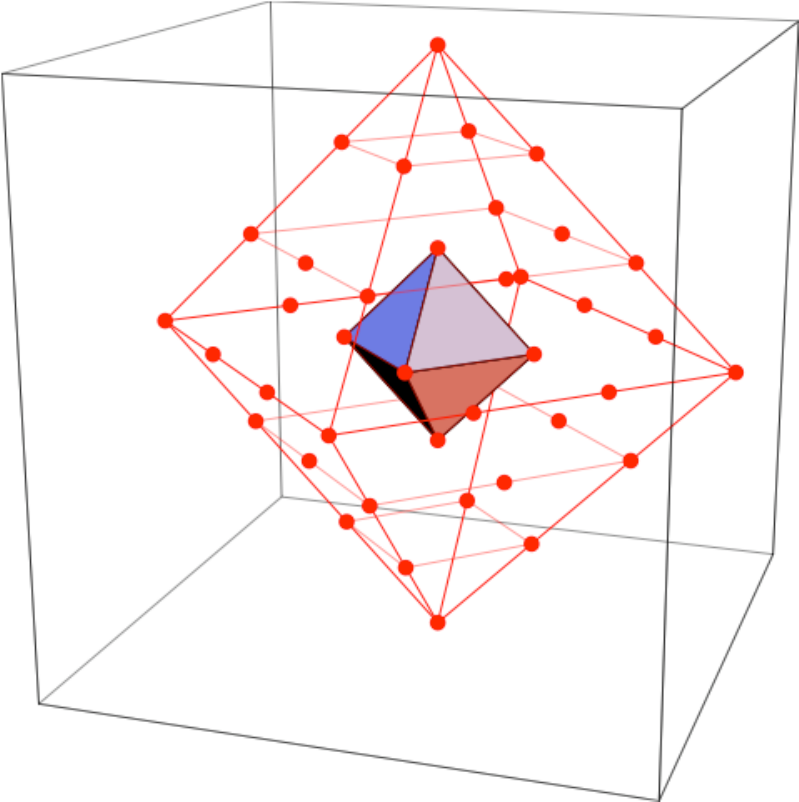}}
\caption{The points represent the states of two IUR's of
$so(6)$ with $q=1$ (left) and $q=3$
(right). The points corresponding to $q=3$ include those of $q=1$   (the inner octahedron) which are double degenerated.} \label{fig3}
\end{figure}

These representations depicted in the parameter space correspond to  octahedrons, that contain iso-energy representations of $su(3)$  labeled by $(m,n)$ such that $m+n=q$ is f\/ixed. In Fig.~\ref{fig3} we represent two IUR's of
$so(6)$ characterized by $q=1$  and $q=3$, respectively.
The  6 $(q=1)$-eigenstates have  energy
$E=\frac52\cdot\frac72$ and the  50 $(q=3)$-eigenstates share
energy $E=\frac72\cdot\frac92$.

\begin{figure}[t]
\centerline{\includegraphics[scale=0.67]{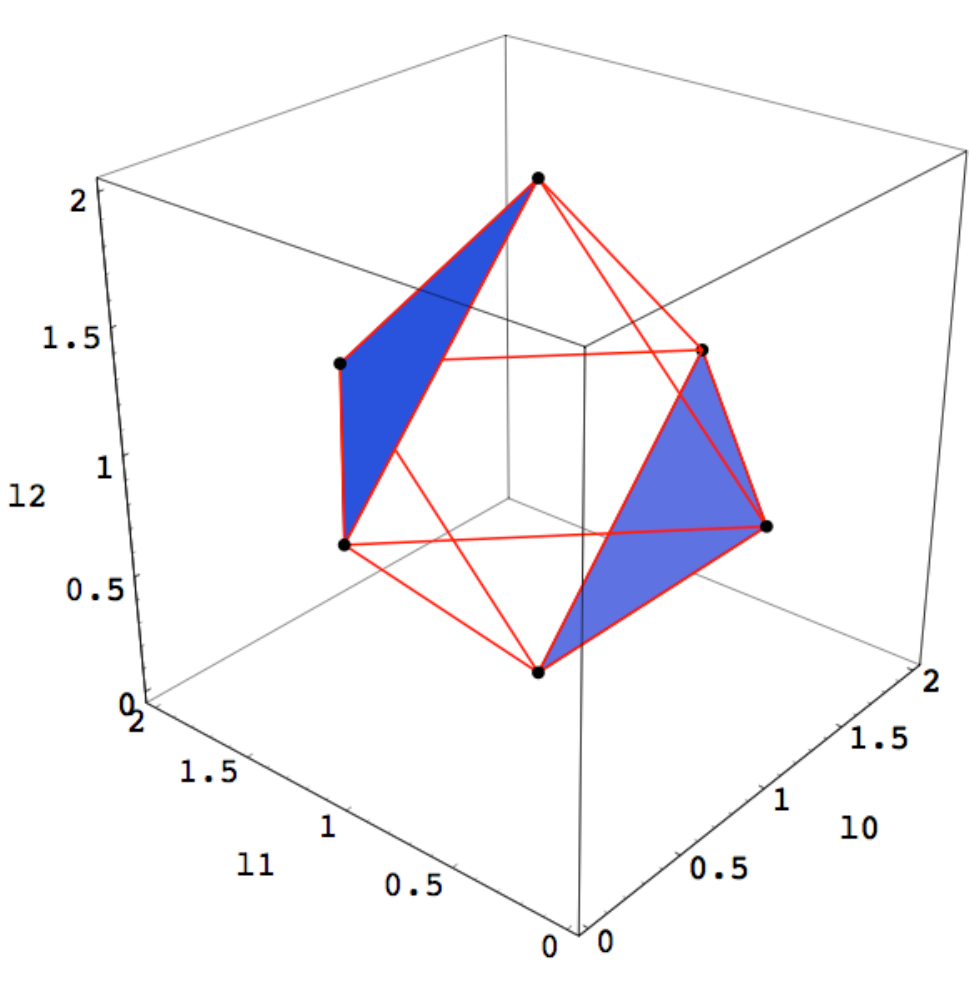}\hskip1.5cm
\includegraphics[scale=0.67]{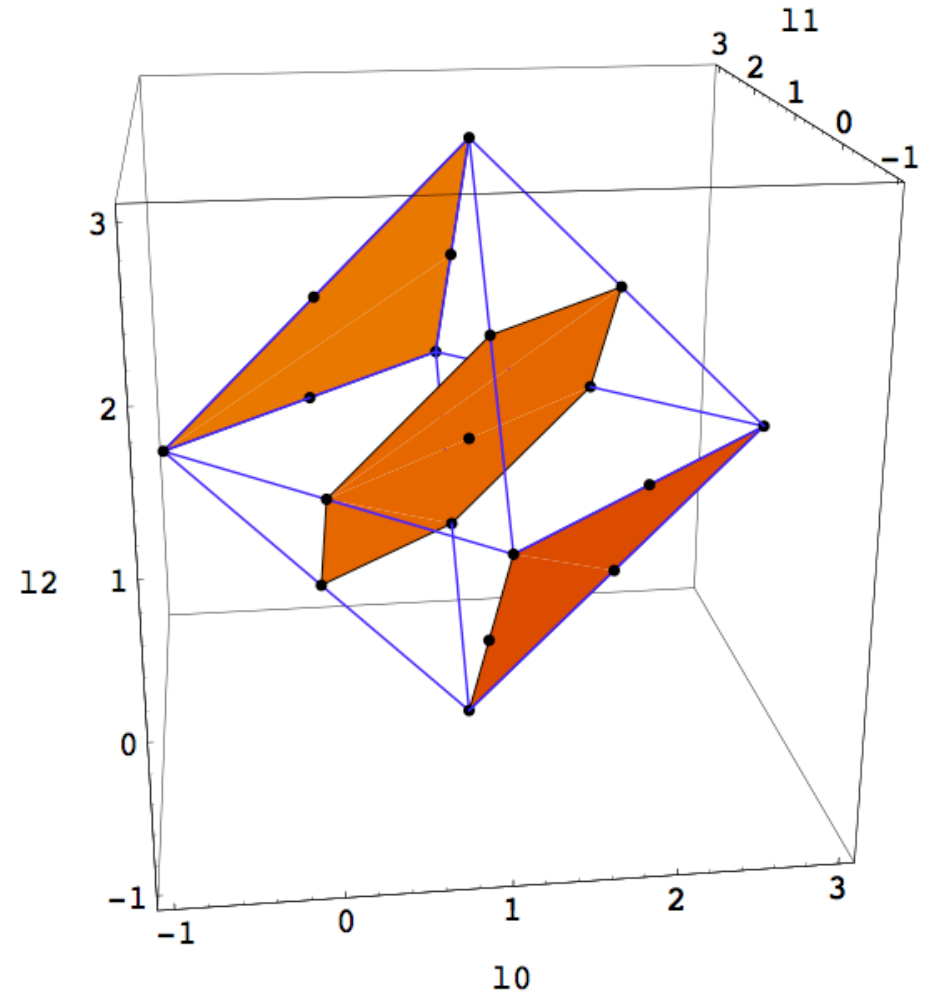}}
\caption{(Left)  $q=1$ IUR of $so(6)$ where the triangular opposite faces correspond to two IUR's of $su(3)$.  (Right) Points of a  $q=2$ IUR of $so(6)$ that are associated to three IUR's of $su(3)$.}
 \label{fig4}
\end{figure}

In Fig.~\ref{fig4} we show how the IUR's of $so(6)$ corresponding to $q=1$ and $q=2$  include IUR's  of~$su(3)$. Thus, in the case $q=1$ the $(m,n)$-IUR's involved are $(1,0)$ and  $(0,1)$: the two parallel exterior faces of the octahedron faces that lie on the plane characterized by $m-n$. In the other case $q=2$,  the $su(3)$-IUR's are $(2,0)$,   $(1,1)$ and  $(0,2)$. The f\/irst and the last ones correspond to the opposite parallel faces of the octahedron, the $(1,1)$-IUR is represented in the parallel hexagonal section containing the origin.

\section{Conclusions}\label{section4}

In this section we will enumerate a list of
interesting features coming from the analysis
of the IO's associated to a SHS, for instance,  by means of the example def\/ined on a
two-sheet hyperboloid. Obviously, from the Hamiltonian living in the sphere we can arrive to the same conclusions.

The IO's of a SHS close an algebraic structure, in this case a
non-compact $su(2,1)$ Lie algebra.
By using the ref\/lections operators of the system we can implement these IO's obtaining a~broader algebra: $so(4,2)$.
These IO's lead to {hierarchies
of Hamiltonians} described by points on planes ($su(2,1)$) or
in the 3-dimensional space ($so(4,2)$), corresponding to the rank of the
respective algebra.
This framework of  IO's can be very helpful in the characterization of a~physical system by selecting separable coordinate systems and determining the eigenvalues and building eigenfunctions.

We have shown  the relation of eigenstates
and eigenvalues with unitary representations of the $su(2,1)$ and
$so(4,2)$ Lie algebras.
In particular, we have studied the degeneration
problem as well as the number of bound states.
Here, we remark that such a detailed study for a `non-compact' superintegrable system had not
been realized till now, up to our knowledge.
We have restricted to IUR's, but a wider analysis can be done
for hierarchies associated to representations with a not well
def\/ined unitary character.

The IO's can also be used to f\/ind the second order
integrals of motion for a Hamilto\-nian~$H_{\ell}$ and their algebraic
relations, which is the usual approach to (super)-integrable
systems.
However, we see that it is much easier to deal directly
with the IO's, which are more elementary and simpler, than with
constants of motion. The second and third order constants of motion close a~quadratic algebra.

By means of the IO technique we have recovered the
algebraic structure of the system that was used in the
Marsden--Weinstein reducing procedure
\[
su(p,q)\stackrel{\rm M-W} \longrightarrow
 so(p,q)\stackrel{\rm factoriz.}\longrightarrow u(p,q)
 \stackrel{\rm discrete} \longrightarrow so(2p,2q).
\]
This is a step to conf\/irm  the conjecture by
Grabowski--Landi--Marmo--Vilasi~\cite{marmo94} that ``any completely integrable system should arise as reduction of a simpler one
(associated for instance to a simple Lie algebra)''.

Our program in the near future is the application of this method
to wider situations.
For example, to be useful when dealing with other SHS, but not necessarily maximally integrable,
or not having a system of separable variable but still allowing algebraic methods.

\subsection*{Acknowledgments}
This work has been partially supported by DGES of the
Ministerio de Educaci\'on y Ciencia of Spain under Project   FIS2005-03989
and Junta de Castilla y Le\'on (Spain) (Project GR224).

\pdfbookmark[1]{References}{ref}
\LastPageEnding

\end{document}